\documentclass[preprintnumbers, floatfix, preprintnumbers, letterpaper, superscriptaddress,nofootinbib, onecolumn]{revtex4}
\pdfoutput=1
\usepackage{graphicx}
\usepackage{amsmath}
\usepackage{amssymb}
\usepackage{subfigure}
\usepackage{hyperref}
\usepackage{url}
\usepackage{xcolor}
\usepackage{color}
\usepackage{mathrsfs}
\usepackage{calrsfs}
\usepackage{amsfonts}
\usepackage{latexsym}
\usepackage{ragged2e}
\usepackage{textcomp}
\usepackage{phaistos}
\usepackage{epstopdf}
\usepackage{ulem}

\makeatletter
\renewcommand\@makefnmark{\hbox{\@textsuperscript{\normalfont\color{purple}\@thefnmark}}}
\renewcommand\@makefntext[1]{%
  \parindent 1em\noindent
            \hb@xt@1.8em{%
                \hss\@textsuperscript{\normalfont\@thefnmark}}#1}

\makeatother
\usepackage{caption}
\DeclareCaptionJustification{justified}{\leftskip=0pt \rightskip=0pt \parfillskip=0pt plus 1fil}
\definecolor{vividviolet}{rgb}{0.62, 0.0, 1.0}
\definecolor{amaranth}{rgb}{0.9, 0.17, 0.31}
\definecolor{palatinateblue}{rgb}{0.15, 0.23, 0.89}
\definecolor{brightpink}{rgb}{1.0, 0.0, 0.5}
\definecolor{cornflowerblue}{rgb}{0.39, 0.58, 0.93}
\definecolor{deepcarminepink}{rgb}{0.94, 0.19, 0.22}
\definecolor{radicalred}{rgb}{1.0, 0.21, 0.37}

\hypersetup{ linktoc=all,
    colorlinks, linkcolor={palatinateblue},
    citecolor={brightpink}, urlcolor={amaranth}
}

\graphicspath{{Images/}}

\graphicspath{{Images/}}

\makeatletter
\def\@fnsymbol#1{\ensuremath{\ifcase#1\or $\PHplaneTree$ \or $\textleaf$ 
\else\@ctrerr\fi}}%

\makeatother

\def\sideremark#1{\ifvmode\leavevmode\fi\vadjust{\vbox to0pt{\vss
 \hbox to 0pt{\hskip\hsize\hskip1em
 \vbox{\hsize1.5cm\tiny\raggedright\pretolerance10000
 \noindent #1\hfill}\hss}\vbox to8pt{\vfil}\vss}}}%
\def\sideremark#1{\ifvmode\leavevmode\fi\vadjust{\vbox to0pt{\vss
 \hbox to 0pt{\hskip\hsize\hskip1em
 \vbox{\hsize1.3cm\tiny\raggedright\pretolerance10000
 \noindent #1\hfill}\hss}\vbox to8pt{\vfil}\vss}}}%
\begin{document}

\title{Non-Abelian Wormholes Threaded by a Yang-Mills-Higgs Field beyond the BPS Limit}

\author{Xiao Yan \surname{Chew}}
\email{xychew998@gmail.com}
\affiliation{Department of Physics Education, Pusan National University, Busan 46241, Republic of Korea}
\affiliation{Research Center for Dielectric and Advanced Matter Physics, Pusan National University, Busan 46241, Republic of Korea}

\author{Kok-Geng \surname{Lim}}
\email{K.G.Lim@soton.ac.uk}
\affiliation{University of Southampton Malaysia, 79200 Iskandar Puteri, Johor, Malaysia}

\begin{abstract}
We construct numerically the symmetric non-Abelian wormholes which are supported by a phantom field in the Einstein-Yang-Mills-Higgs theory beyond Bogomol'nyi-Prasad-Sommerfield (BPS) limit where the Higgs self-interaction constant $\lambda$ is non-vanishing. Analogous to the BPS limit, the probe limit is the Yang-Mills-Higgs field in the background of the Ellis wormhole when the gravity is switched off. In the presence of gravity, the wormhole solutions possess the Yang-Mills-Higgs hair where families of hairy wormholes solutions emerge from the Ellis wormhole when the gravitational coupling constant increases. In contrast to the BPS limit, the properties of wormholes change drastically when the gravitational strength approaches a critical value for a fixed $\lambda$. The hairy wormholes possess two types of double throat configurations. The first type is wormholes that develop the double throat when the gravitational strength almost approaches the critical value for lower $\lambda$, whereas the second type is wormholes that exhibit the double throat for a certain range of gravitational strength for higher $\lambda$. These two types of double throat configurations can coexist for a certain range of $\lambda$ where it is a transition process for first type double throat disappears gradually and the second type double throat becomes dominant.

\begin{center}

\end{center}
\end{abstract}

\maketitle

\section{Introduction}

The solitonic magnetic monopole with finite energy is an inevitable outcome of the non-Abelian SU(2) Yang-Mills-Higgs (YMH) theory upon spontaneously broken by the Higgs field to a residual symmetry U(1) \cite{rossi1982exact,Shnir:2005vvi}. The non-Abelian magnetic monopole discovered by 't Hooft and Polyakov is a three-dimensional topological soliton \cite{t1974magnetic,polyakov1974particle}, whereby its topological charge corresponds to the magnetic charge. Exact monopole \cite{prasad1975exact} and axially symmetric multimonopole solutions \cite{rebbi1980multimonopole,forgacs1981exact} are found in the Bogomol'nyi-Prasad-Sommerfield (BPS) limit \cite{bogomol1976stability} where the Higgs potential vanishes, and the Higgs field becomes massless. These BPS solutions satisfy the first-order Bogomol'nyi equation and their mass saturated at the lower Bogomol'nyi bound \cite{bogomol1976stability}. Beyond the BPS limit, when the Higgs potential is non-vanishing and the Higgs field becomes massive, the non-BPS solutions of monopole \cite{Bogomolnyi1976calculation,bais1976integral,teh2010generalized} and multimonopole \cite{kleihaus1998interaction} no longer fulfil the Bogomol'nyi equation but only can be numerically obtained by solving the second-order Euler-Lagrange equations of YMH theory. Likewise, the monopole-antimonopole pair \cite{kleihaus1999monopole}, vortex ring and monopole-antimonopole chain \cite{kleihaus2003monopole,kleihaus2004monopole} also do not satisfy the Bogomol'nyi equation and they are the non-BPS, saddle-points solutions in flat space which possess finite energy.

The coupling of the YMH model with gravity or known as Einstein-Yang-Mills-Higgs (EYMH) theory gives rise to a branch of gravitating monopole solutions \cite{lee1992black,Breitenlohner:1991aa,Breitenlohner:1994di,Brihaye:1998cm,Lue:1999zp,Brihaye:1999nn,Brihaye:1999kt, Brihaye:1999zt,Brihaye:2002pc}. The mass of the gravitating monopole strictly decreases as the gravitational strength increases.  The solutions of gravitating monopole, which includes the static axially symmetric gravitating monopole solutions \cite{Hartmann:2000gx,Hartmann:2001ic} end up as an extremal Reissner-Nordstrom black hole when the gravitational strength reaches a critical value. Note that the black holes in the EYMH system are dubbed as the ``black hole within monopole'' \cite{lee1992black,Brihaye:1998cm} because they possess the non-abelian gauge field outside the event horizon, which is a counterexample to the ``no hair'' conjecture for black holes. Similar to the YMH theory, the EYMH theory also reported various gravitating soliton solutions such as gravitating monopole and antimonopole pairs \cite{Kleihaus:2000hx,Kleihaus:2000kv,Hartmann:2001ic,VanderBij:2001nm,Paturyan:2003hz,Paturyan:2004ps,Kleihaus:2004gm} and vortex rings \cite{Kleihaus:2004fh,Kleihaus:2005fs}.

Besides, some wormhole-like structures in EYMH theory have been reported in \cite{Hajicek_1983a,Hajicek_1983,1987GReGr..19..739D}. Recently the authors have constructed numerically the solutions of symmetric wormholes in the EYMH system for the BPS limit \cite{Chew:2020svi}. The throat of the wormholes is supported by a phantom field which can violate the null energy condition in order to prevent the collapse of the throat for the construction of traversable wormholes in general relativity. A phantom field is a real-valued scalar field that has an opposite sign for the kinetic term in the Lagrangian. It has been used to construct the classic example of traversable wormhole, which is known as the Ellis wormhole \cite{Ellis:1973yv,Ellis:1979bh,Bronnikov:1973fh}. However,  the Ellis wormhole possesses the unstable radial modes \cite{Gonzalez:2008wd,Gonzalez:2008xk,Torii:2013xba}. The Ellis wormhole has been generalized to the higher-dimensional case \cite{Torii:2013xba}, the slowly rotating case with perturbative method \cite{Kashargin:2007mm,Kashargin:2008pk}, the rapidly rotating case in four dimensions \cite{Kleihaus:2014dla,Chew:2016epf} and five dimensions with equal angular momenta \cite{Dzhunushaliev:2013jja}, in the modified gravity, e.g., the scalar-tensor theory \cite{Chew:2018vjp} and {\it f}(R) gravity \cite{Karakasis:2021tqx}. Recently, the Ellis wormhole has been considered in the bouncing universe \cite{Huang:2020qmn} and generalized in asymptotic anti-de Sitter \cite{Blazquez-Salcedo:2020nsa}.

In the BPS limit, the EYMH wormholes admit the Ellis wormhole as the trivial solution, which is analogous to Schwarzschild black hole when the gauge fields vanish \cite{Chew:2020svi}. Analogous to particle-like solutions of EYMH, the corresponding wormholes possess a probe limit where only the gauge field is present in the background solution of Ellis wormhole when the gravitational field strength vanishes. When the gravity present, a branch of hairy wormholes emerge from the Ellis wormhole where the wormholes gain the mass and its phantom scalar charge increases. The wormholes possess a double-throat configuration when the gravitational field strength exceeds a value.

In addition, another non-abelian wormhole in Einstein-Yang-Mills (EYM) and phantom field system has been obtained numerically \cite{hauser2014hairy}. Similarly, these hairy wormholes solutions also possess a probe limit without the presence of gravity. These hairy wormholes solutions possess a sequence of solutions, which are labelled by the node number $k$ of the gauge field function. They are analogous to the Bartnik-McKinnon solution, which is the regular and spherically symmetric solutions of the EYM system \cite{bartnik1988particlelike}.

On the other hand, since wormholes can be a candidate for the black hole mimicker, then several astrophysical signatures of wormholes have been proposed to search for their existence in near future, for example, the shadow \cite{Nedkova:2013msa,Gyulchev:2018fmd,Amir:2018szm,Narzilloev:2021ygl,Benavides-Gallego:2021lqn,Jusufi:2021lei,Bambi:2021qfo}, the gravitational lensing \cite{Abe:2010ap,Toki:2011zu,Takahashi:2013jqa,Cramer:1994qj,Perlick:2003vg,Tsukamoto:2012xs,Bambi:2013nla}, the accretion disk around the wormhole \cite{Zhou:2016koy,Deligianni:2021ecz,Deligianni:2021hwt}, and the ringdown phase in the emission of gravitational waves \cite{Blazquez-Salcedo:2018ipc}.

Since EYMH wormholes for the BPS limit in our previous work \cite{Chew:2020svi} give rise to new and interesting phenomena due to the presence of the non-Abelian field. In this paper, our motivation is to continue our investigation by numerically obtaining the symmetric wormhole solutions in EYMH with the presence of Higgs potential and analyze their properties. Thus our paper is organized as follows. In section II, we briefly introduce the EYMH theory and present the equation of motions. Subsequently, we introduce the geometrical properties of the wormhole. We then obtain the ordinary differential equations (ODEs) from the equation of motions and discuss the global charges of the wormholes. In section III, we exhibit and discuss our numerical results. In section IV, we conclude our research work and discuss the possible outlook from this present work.

\section{Theoretical Framework}

\subsection{Einstein-Yang-Mills-Higgs Theory}

The Einstein gravity couples with a phantom field $\psi$ and a gauge field $A_\mu$ in SU(2) YMH theory in the Einstein-Hilbert action,
\begin{equation} \label{EHaction}
 S =  \int d^4 x \sqrt{-g}  \left[  \frac{R}{16 \pi G}  + \mathcal{L}_{\text{ph}} + \mathcal{L}_{\text{YMH}} \right]  \,,
\end{equation}
where the Lagrangian of phantom field and YMH \cite{Hartmann:2001ic} are respectively, given by 
\begin{equation}
  \mathcal{L}_{\text{ph}} = \frac{1}{2} \partial_\mu \psi \partial^\mu \psi \,, \quad \mathcal{L}_{\text{YMH}} =  - \frac{1}{2} \text{Tr}\left( F_{\mu \nu} F^{\mu \nu} \right) - \frac{1}{4} \text{Tr} \left(  D_{\mu} \Phi D^{\mu} \Phi  \right) - \frac{\lambda}{8}  \text{Tr}  (\Phi^2-\upsilon^2)^2   \,,
\end{equation}
where $\lambda$ is the Higgs field self-interaction constant and $\upsilon$ is the vacuum expectation value of the Higgs field. The covariant derivative of the Higgs field and the gauge field strength tensor are given respectively by 
\begin{equation}
F_{\mu\nu} = \partial_{\mu}A_\nu - \partial_{\nu}A_\mu + i \left[ A_{\mu}, A_\nu \right] \,, \quad
D_{\mu} \Phi = \partial_{\mu} \Phi + i \left[ A_{\mu}, \Phi \right] \,, 
\label{1.1.2}
\end{equation}
where $A_\mu = \frac{1}{2} \tau^a A_\mu^a$ and $\Phi=\phi^a \tau^a$ with $\tau^a$ is the Pauli matrices. Since we only construct spherically symmetric wormholes, then we also employ the spherically symmetric Ansatz in a purely magnetic gauge field $(A_t=0)$ for the gauge and Higgs field \cite{Brihaye:1998cm}, 
\begin{equation} \label{gaugefield}
 A_{\mu} dx^\mu = \frac{1-K(\eta)}{2} \left(  \tau_{\varphi}  d \theta -   \tau_\theta  \sin \theta d\varphi  \right) \,, \quad \Phi =  H(\eta) \tau_\eta  \,.
\end{equation}

The action is varied with respect to the metric $g_{\mu \nu}$, which yields the Einstein equation,
\begin{equation} \label{einstein_eqn}
 R_{\mu \nu} - \frac{1}{2} g_{\mu \nu} R =  \beta \left( T_{\mu \nu}^{\text{ph}} + T_{\mu \nu}^{\text{YMH}} \right) \,,
\end{equation}
where $\beta=8 \pi G$, the stress-energy tensor for phantom field $T_{\mu \nu}^{\text{ph}}$ and YMH $T_{\mu \nu}^{\text{YMH}}$ are respectively,  given by
\begin{align}
 T_{\mu \nu}^{\text{ph}} &= \frac{1}{2} g_{\mu \nu} \partial_\alpha \psi \partial^\alpha \psi -  \partial_\mu \psi \partial_\nu \psi \,, \\
 T_{\mu \nu}^{\text{YMH}} &= \text{Tr} \left( \frac{1}{2} D_\mu \Phi D_\nu \Phi - \frac{1}{4} g_{\mu \nu} D_\alpha \Phi D^\alpha \Phi      \right) + 2 \text{Tr} \left(  g^{\alpha \beta} F_{\mu \alpha} F_{\nu \beta} - \frac{1}{4} g_{\mu \nu} F_{\alpha \beta} F^{\alpha \beta}   \right) - \frac{\lambda}{8} g_{\mu \nu} \text{Tr}  (\Phi^2-\upsilon^2)^2\,.
\end{align}
The equations of motion for the matter fields are
\begin{equation} \label{matter_eqn}
  \frac{1}{\sqrt{-g}}  \partial_\mu \left(  \sqrt{-g} \partial^\mu \psi  \right)  = 0 \,,  \quad D_\mu  F^{\mu \nu}=\frac{i }{4} \left[ \Phi, D^{\nu} \Phi   \right]  \,, \quad D_\mu   D^\mu \Phi = \lambda (\Phi^2-\eta^2) \Phi  \,.
\end{equation}

\subsection{The Geometrical Structure of Wormhole}

A globally regular wormhole spacetime can be constructed by employing the quasi-isotropic line element, 
\begin{equation}  \label{line_element}
 ds^2 = - F_0(\eta) dt^2 + F_1(\eta) \left[  d \eta^2  +  h(\eta) \left( d \theta^2+\sin^2 \theta d\phi^2 \right) \right]  \,,
\end{equation}
where $h(\eta)=\eta^2+\eta_0^2$ with $\eta_0$ as the throat parameter. The wormhole spacetime possesses two asymptotically flat regions in the limit $\eta \rightarrow \pm \infty$. Here we define the function $R(\eta)^2$ as the circumferential radius of the wormhole,
\begin{equation}
 R(\eta)^2 = F_1 h \,.
\end{equation} 
Note that $R(\eta)$ should not contain zero for a globally regular wormhole solution. When $R$ contains a local minimum, which is known as the throat of wormhole $\eta_{\text{th}}$, then the wormhole possesses a minimal surface area at the throat,  $A_{\text{th}}=4 \pi R(\eta_{\text{th}})^2$. However, if $R$ contains a local maximum, then it is an equator of the wormhole $\eta_{\text{eq}}$, which corresponds to the maximal surface area of the wormhole, $A_{\text{eq}}=4 \pi R(\eta_{\text{eq}})^2$. The equator of the wormhole is normally sandwiched between two throats. 

The wormholes that possess either a throat or an equator can be determined by the following conditions, respectively
\begin{eqnarray}
1. \quad  R'(\eta_{\text{th}}) = 0 \,,  \ &\text{and}& \  R''(\eta_{\text{th}}) > 0 \,, \\
2. \quad  R'(\eta_{\text{eq}}) = 0 \,,  \ &\text{and}& \ R''(\eta_{\text{eq}}) < 0 \,.
\end{eqnarray}
When $R'(\eta_\text{crit})=R''(\eta_\text{crit})=0$, the geometry of wormhole is in a transition state because the circumferential radius forms a turning point at some value of the radial coordinate $\eta_\text{crit}$, such that the double throat and the equator can simultaneously exist, this also implies that there is a transition can occur from the single throat configuration to the double throat configuration \cite{Dzhunushaliev:2014bya,Hoffmann:2017jfs}.

In this paper, we only consider the metric functions symmetric w.r.t. the coordinate $\eta=0$, thus the circumferential radius of the wormhole at $\eta=0$ is assumed to be either a throat or an equator, which implies $R$ should have an extremum at $\eta=0$ by demanding
\begin{equation}
  R'(0) = 0 \quad \Rightarrow \quad \frac{\left( h F'_1 + 2 \eta F_1 \right)}{2 R}   \Bigg|_{\eta=0} = 0\,,
\end{equation}
where we have to set $F'_1(0)=0$. In particular, if a wormhole only contains a single throat, then the throat must be located at $\eta=0$ with the minimal surface area $A_{\text{th}}= 4 \pi R(0)^2=4 \pi F_1(0) \eta_0^2$.

\subsection{Ordinary Differential Equations (ODEs)}

A set of second-order and nonlinear ODEs is obtained for the metric functions and gauge fields by substituting Eqs. \eqref{line_element} and \eqref{gaugefield} into the Einstein equation Eq. \eqref{einstein_eqn} and equations of motion for the gauge fields Eq. \eqref{matter_eqn}, 
\begin{align}
& F''_1 + \frac{2 \eta}{h} F'_1 -  \frac{3 F'^2_1}{4 F_1} +  \frac{\eta_0^2 F_1 }{h^2}    \nonumber \\
&\qquad \qquad \qquad =  \beta \frac{F_1}{2} \psi'^2 - \beta \frac{(K^2-1)^2 + 2 h K'^2 + h^2 F_1 H'^2 + 2 h F_1 H^2 K^2}{ 2 h^2 } - \beta \frac{\lambda}{4} F_1^2 \left( H^2-\upsilon^2 \right)^2     \,, \label{ode1} \\
& \left(  \frac{F'_1}{2 F_1} + \frac{\eta}{h} \right) \frac{F'_0}{F_0}   + \frac{F'^2_1}{4 F^2_1} +  \frac{\eta}{h F_1} F'_1 -   \frac{\eta_0^2}{h^2 }  \nonumber \\
& \qquad \qquad \qquad  = - \frac{ \beta}{2} \psi'^2  + \beta  \frac{ - (K^2-1)^2 + 2 h K'^2 + h^2 F_1 H'^2 - 2 h F_1 H^2 K^2}{2 h^2 F_1}  - \beta \frac{\lambda}{4} F_1 (H^2-\upsilon^2)^2  \,,  \label{ode2}  \\
&F''_0 + \left( - \frac{F'_0}{2 F_0} + \frac{\eta}{h} \right) F'_0 +  \frac{F_0}{F_1} F''_1 +  \left( - \frac{F'_1}{F_1} + \frac{\eta}{h} \right) \frac{F_0 F'_1}{F_1}   + \frac{2 F_0 \eta_0^2}{h^2}    \nonumber \\
& \qquad  \qquad \qquad =  \beta F_0 \psi'^2  -  F_0 \left[ \frac{-(K^2-1)^2 + h^2 F_1 H'^2}{ h^2 F_1}  \right] - \beta \frac{\lambda}{2} F_0 F_1  (H^2-\upsilon^2)^2   \,. \label{ode3}    \, \\
K'' &+ \frac{1}{2} \left(  \frac{F'_0}{F_0} - \frac{F'_1}{F_1}    \right) K' = \frac{K (K^2-1+ h F_1 H^2)}{h}  \,,  \label{odeYM}  \\
 H'' &+ \frac{1}{2} \left(  \frac{F'_0}{F_0} + \frac{F'_1}{F_1} + \frac{4 \eta}{h}   \right) H' - \frac{2 K^2}{h} H = \lambda F_1 ( H^2-\upsilon^2  ) H  \,. \label{odeHiggs} 
\end{align}
where the prime denotes the derivative of the functions w.r.t. the radial coordinate $\eta$. The equation of motion Eq. \eqref{matter_eqn} for the phantom field yields a first-order integral, $\psi' =  D/ ( h \sqrt{F_0 F_1})$
where $D$ is the scalar charge of the phantom field. Thus, the term $\psi'^2$ in Eqs.~\eqref{ode1}-\eqref{ode3} can be replaced by $\psi' =  D/ ( h \sqrt{F_0 F_1})$. We subtract Eqs.~\eqref{ode1} and \eqref{ode3} with Eq.~\eqref{ode2} to obtain the final set of ODEs,
\begin{align}
 F''_0 &=   \frac{F'_0}{2} \left( \frac{F'_0}{F_0} - \frac{F'_1}{F_1} -  \frac{4 \eta}{h} \right) +  \beta F_0 \frac{(K^2-1)^2 + 2 h  K'^2}{h^2 F_1} - \frac{1}{2}  \beta F_0 F_1 \lambda  ( H^2-\upsilon^2  )^2  \,,  \label{Minset1} \\
 F''_1 &= \frac{F'^2_1}{2 F_1} - \frac{3 \eta F'_1}{h} - \left(  \frac{F'_1}{2} + \frac{F_1}{h}  \right)  \frac{F'_0}{F_0}  -  \frac{\beta  \left[  (K^2-1)^2 + 2 h F_1 H^2 K^2  \right]}{ h^2} - \frac{\beta \lambda}{2} F_1^2  ( H^2-\upsilon^2  )^2  \,,  \label{Minset2}\\
 K'' &= \left(  \frac{F'_1}{F_1} - \frac{F'_0}{F_0}   \right)   \frac{K'}{2} + \frac{K \left( K^2-1+ h F_1 H^2 \right)}{h} \,,  \label{Minset3}  \\
 H'' &= -  \frac{1}{2} \left( \frac{F'_0}{F_0} + \frac{F'_1}{F_1} + \frac{4 \eta}{h}\right) H' + \frac{2 H K^2}{h} + \lambda F_1 ( H^2-\upsilon^2  ) H \,,  \label{Minset4}
\end{align}
 with Eq. \eqref{ode2} is expressed as 
\begin{align}
 D^2 &=  \frac{2 h^2 F_0 F_1}{\beta} \left[   - \left(  \frac{F'_1}{2 F_1} + \frac{\eta}{h} \right) \frac{F'_0}{F_0}   - \frac{F'^2_1}{4 F^2_1} -  \frac{\eta}{h F_1} F'_1 +  \frac{\eta_0^2}{h^2 }  \nonumber  \right. \\
& \qquad \qquad \left. + \beta  \frac{ - (K^2-1)^2 + 2 h K'^2 + h^2 F_1 H'^2 - 2 h F_1 H^2 K^2}{2 h^2 F_1}  - \beta \frac{\lambda}{4} F_1 (H^2-\upsilon^2)^2   \right] \,. \label{Constr}
\end{align}
to monitor the quality of the numerical solutions with the condition $D^2=const$. 

Since we only consider the wormhole solutions with the metric functions symmetric w.r.t. $\eta=0$, thus we solve Eqs. \eqref{Minset1}-\eqref{Minset4} numerically from $\eta=0$ to the infinity by Colsys and Matlab package bvp4c. Both packages solve the boundary value problems for systems of nonlinear coupled ODEs, Colsys is based on the Newton-Raphson method with adaptive mesh refinement and error estimation for the solutions \cite{colsys} whereas  bvp4c implements the three-stage Lobatto IIIa collocation formula \cite{kierzenka2001bvp}. In the numerics, we impose the eight boundary conditions at $\eta=0$ and $\eta=\infty$. First, we let the the metric functions possess the extremum at $\eta=0$ by requiring the first order derivative of the metric functions vanish at the throat, $F'_0 (0) = F'_1 (0) = 0$. The metric functions approach Minkowski spacetime at the infinity, $F_0 (\infty) = F_1 (\infty) = 1$. We impose the following boundary conditions $ K(0)=1$, $H(0) = 0$, $K(\infty) = 0$, $H(\infty)=1$ for the gauge fields. Furthermore, we compactify the radial coordinate $\eta$ by $\eta= \eta_0 \tan \left( \pi x/2 \right)$ in the numerics. We also scale some parameters by the throat parameter $\eta_0$,
\begin{equation}
 \eta \rightarrow \eta_0 \eta \,, \quad  h \rightarrow \eta_0^2 h \,, \quad  \beta \rightarrow  \eta^2_0 \beta \,, \quad H \rightarrow \frac{ H}{\eta_0} \,, \quad \upsilon \rightarrow \frac{\upsilon}{\eta_0} \,. 
\end{equation}

Besides, the second-order derivative of $R$ at $\eta=0$ is given by 
\begin{equation}
R''(0)  =  \frac{2 F_1+h F''_1}{2 R} \Bigg|_{\eta=0} = \frac{2 F_1-\frac{1}{2} F^2_1 \beta \lambda \eta^2_0 \upsilon^4}{2 R}  \,,
\end{equation}
where we have used $F''_1(0)$ from the ODEs. In the BPS limit, $R(0)$ always remains as the throat since $R''(0)>0$.

\subsection{Global Charges}

\begin{figure}[t!]
\centering
\mbox{
(a)
 \includegraphics[angle =-90,scale=0.3]{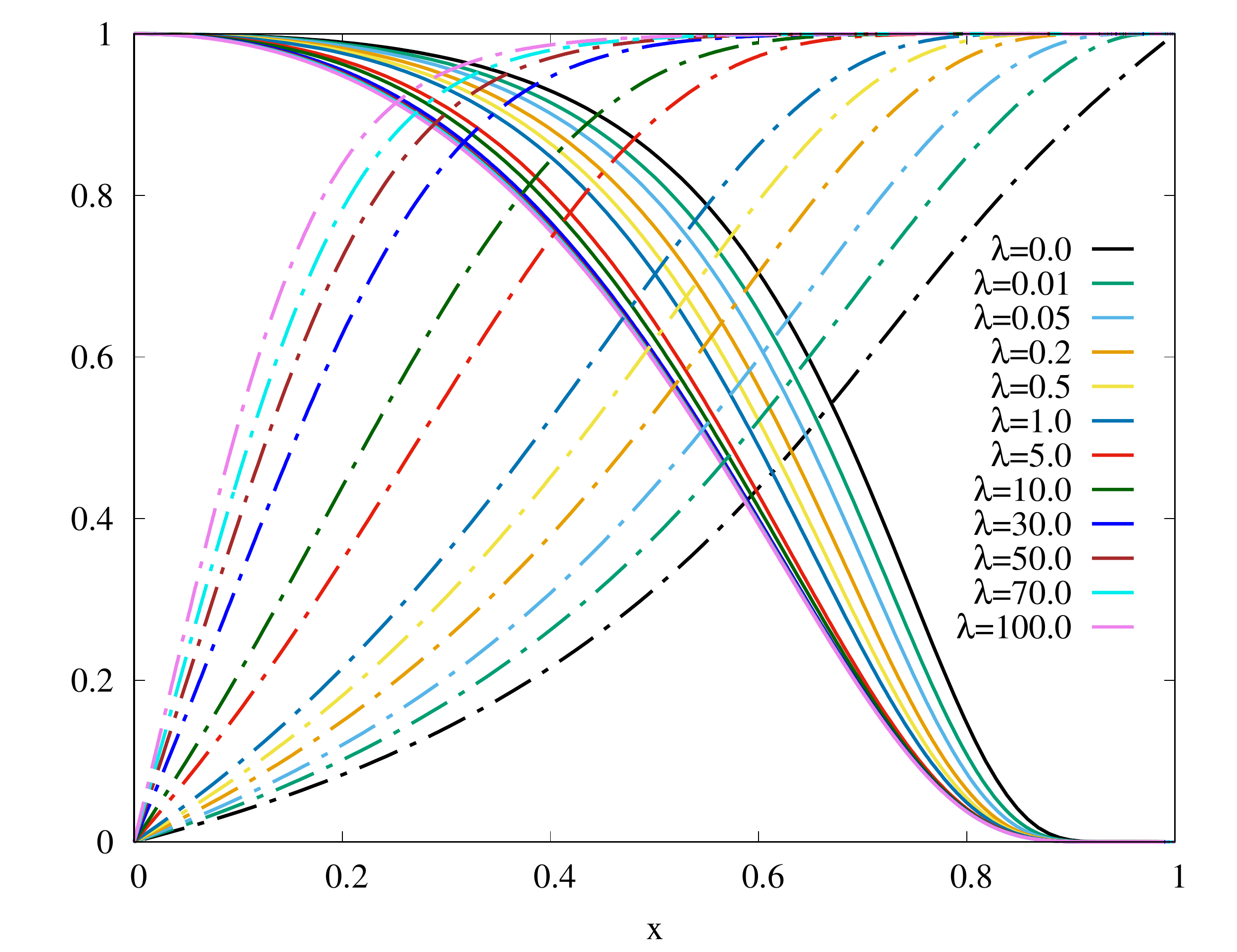}
(b)
 \includegraphics[angle =-90,scale=0.3]{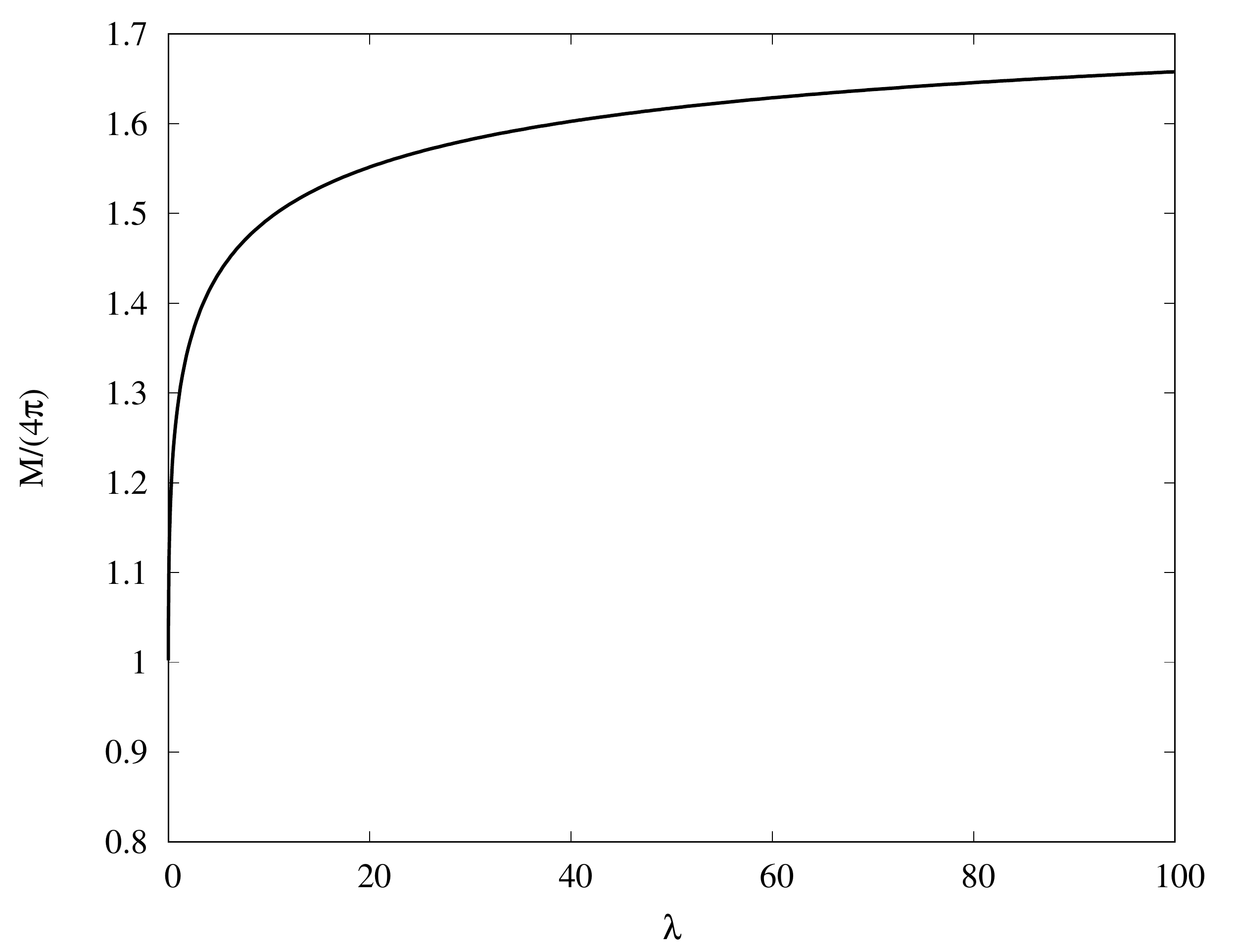}
 }
\mbox{
(c)
 \includegraphics[angle =-90,scale=0.3]{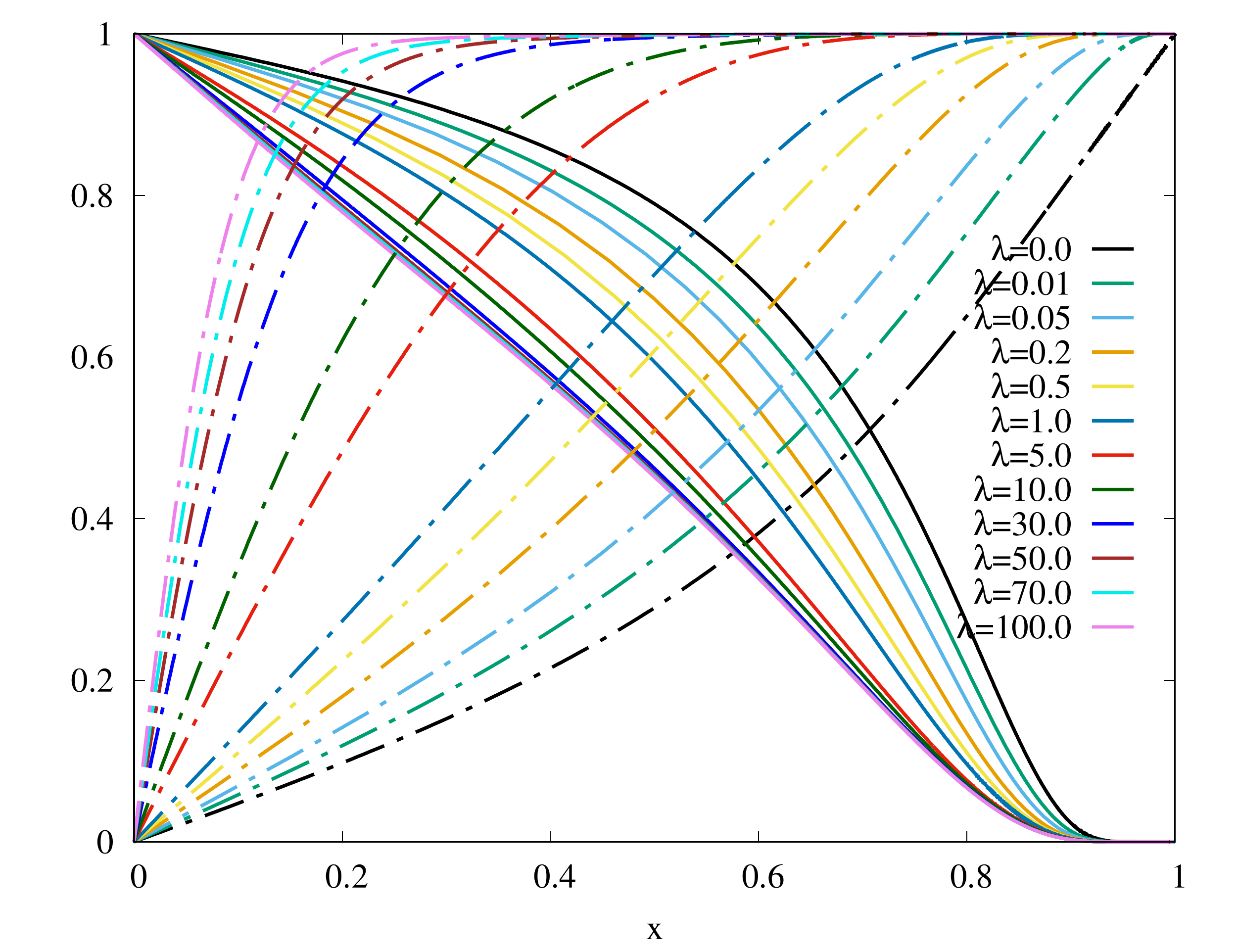}
}
\caption{(a) The solutions of Hooft-Polyakov monopole: $K(x)$ (solid line) and $H(x)$ (dash-dotted line) for several values of $\lambda$ in the compactified coordinate, $x=R/(1+R)$. (b) The mass of Hooft-Polyakov monopole as a function of $\lambda$. (c) The gauge fields $H(x)$ (dash-dotted line) and $K(x)$ (solid line) in the probe limit of hairy wormholes for several values of $\lambda$ in the compactified coordinate $x$.}
\label{plot_HP}
\end{figure}

The mass of wormhole can be read off directly from the asymptotic expansion of the metric at $\eta \rightarrow \infty$,
\begin{equation}
 F_0 \rightarrow 1- \frac{2 G M}{\eta} = 1 - \frac{2 \mu}{\bar{\eta}}  \,,
\end{equation}
where $\mu$ is the mass parameter,
which is given by
\begin{equation}
 \mu = \frac{\beta M}{8 \pi \eta_0} \,.
\end{equation}
One also can obtain the expression for the mass via the Komar integral \cite{hauser2014hairy},
\begin{equation}
 M = M_{\text{th}}+ \frac{1}{4 \pi G} \int_{\Sigma} R_{\mu \nu} \xi^\mu n^\nu dV  = M_{\text{th}}+ \frac{1}{4 \pi G} \int R^0_0 \sqrt{-g}  d^3 x \,,
\end{equation}
where $\Sigma$ is the spacelike hypersurface ($0 \leq \eta < \infty$), $n^\nu $ is a normal vector on $\Sigma$, $\xi^\mu=(1,0,0,0)$ is a timelike Killing vector, $dV$ is the volume element on $\Sigma$.
The term $ M_{\text{th}}$ is the contribution of the throat to the mass, 
\begin{equation}
 M_{\text{th}} = \frac{\kappa A_{\text{th}}}{4 \pi G} \,,
\end{equation}
where $\kappa$ is the surface gravity at the throat, which is given by \cite{Chew:2020svi}
\begin{equation} 
 \kappa = \frac{F'_0}{2 \sqrt{F_0 F_1} } \,. \label{surgra}
\end{equation}
Eq. \eqref{surgra} shows that $\kappa$ and $ M_{\text{th}}$ vanish for symmetric wormholes with only a single throat but remains finite for wormholes with double throat configuration.

The charge of the phantom field is given by $D^2$ and the magnetic charge for the non-abelian gauge fields is given by \cite{Corichi:2000dm,Ashtekar:2000nx}
\begin{equation}
 \mathcal{P}^{\text{YMH}} = \frac{1}{4 \pi} \oint \sqrt{ \sum_i \left(  F^i_{\theta \varphi} \right)^2    } d\theta d\varphi = |P| \,,
\end{equation}
where the integral is evaluated at the spatial infinity, yielding $P=0$ for the hairy wormholes \cite{hauser2014hairy}.

\section{Results and Discussions}

\subsection{Probe Limit}

\begin{figure}[t!]
\centering
\mbox{
(a)
 \includegraphics[angle =-90,scale=0.3]{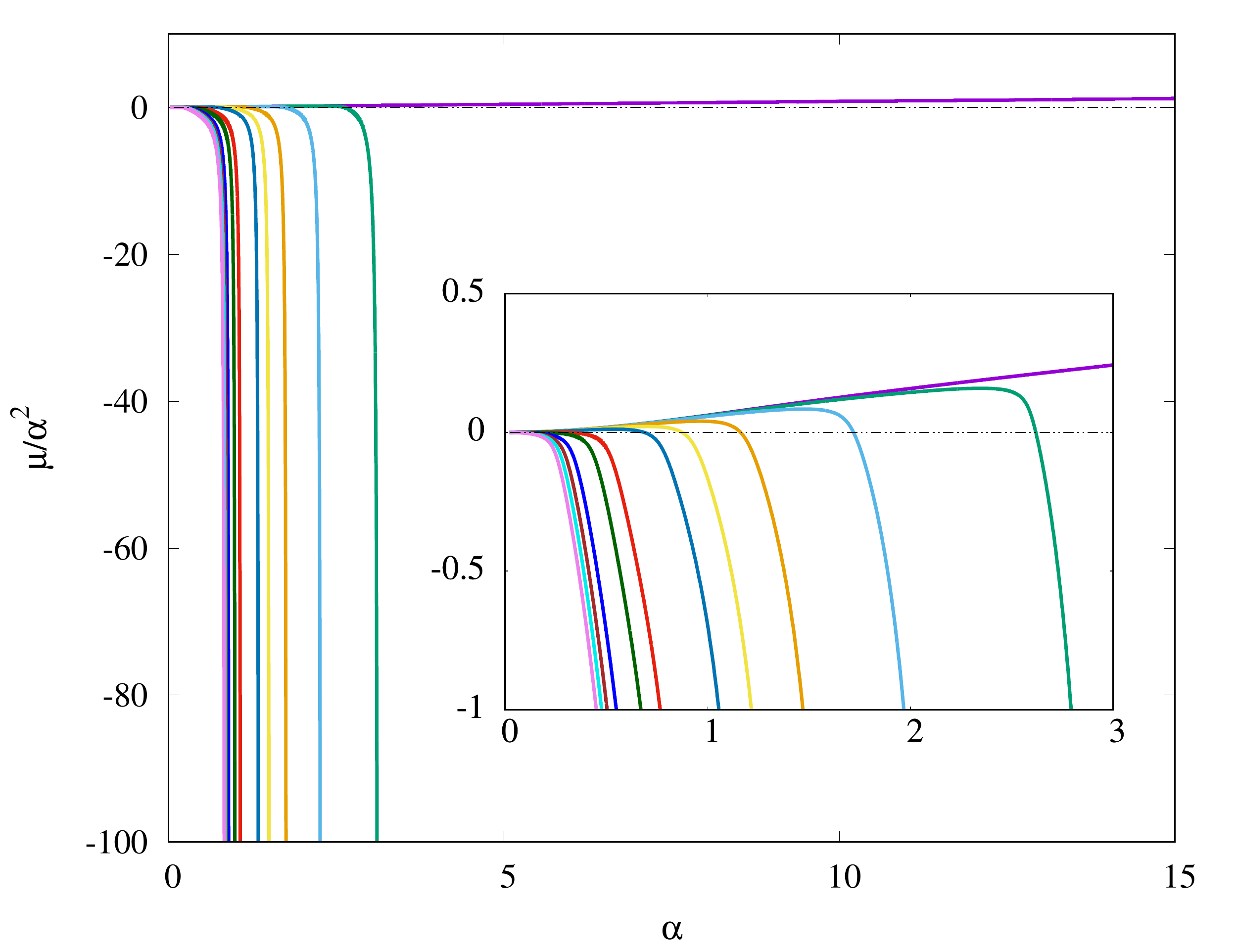}
(b)
 \includegraphics[angle =-90,scale=0.3]{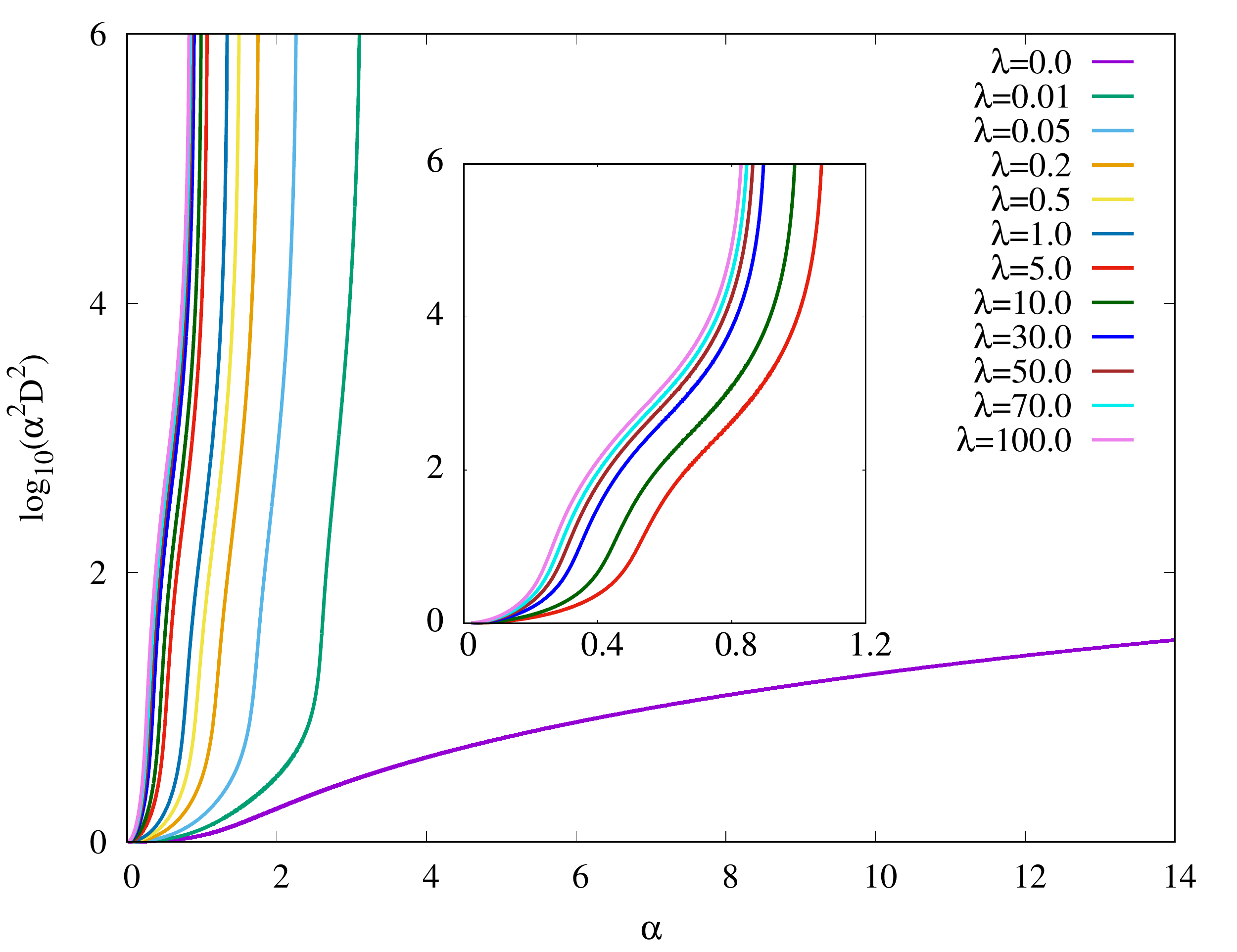}
 }
\caption{The global charges of wormhole solutions for several values of $\lambda$: (a) The scaled mass $\mu/\alpha^2$ versus the scaled gravitational coupling constant $\alpha$. (b) The logarithmic of scaled scalar charge $\log_{10}(\alpha^2 D^2)$ versus the scaled gravitational coupling constant $\alpha$. Note that $\beta=2\alpha^2$.}
\label{plot_prop}
\end{figure}

Here we fix the vacuum expectation value $\upsilon=1$ in all numerics computation. In this part, we discuss the probe limit of hairy wormholes. But first, let us recall that YMH theory possesses a probe limit which is known as Hooft-Polyakov monopole. In the BPS limit, it possesses an exact solution,
\begin{equation}
 K(R) = \frac{R}{\sinh(R)} \,, \quad H(R) = \text{coth}(R)-\frac{1}{R} \,.
\end{equation}
However, the solutions of Hooft-Polyakov monopole beyond the BPS limit can only be obtained numerically which are shown in Fig.\,\ref{plot_HP}(a). The mass of Hooft-Polyakov monopole as shown in Fig.\,\ref{plot_HP}(b) increases monotonically from the unity which corresponds to the BPS limit.

When the gravity is switched off $(\beta=0)$, the metric Eq. \eqref{line_element} in the probe limit of hairy wormholes is the massless Ellis wormhole $(F_0 (\eta) = F_1 (\eta)= 1)$ in which the spacetime is symmetric and has two identical asymptotically flat regions. The phantom field $\psi$ is given exactly by
\begin{equation}
 \psi = \frac{D}{\eta_0} \left[  \arctan \left( \frac{\eta}{\eta_0} \right)  - \frac{\pi}{2} \right] \,.
\end{equation}
Hence, we obtain the pure YMH equations in the background of the Ellis wormhole because the YMH field doesn't contribute to the Einstein equation,
\begin{align}
 K'' &= \frac{K (K^2-1+ h H^2)}{h}  \,, \\
 H'' &= - \frac{2 \eta }{h }  H' + \frac{2 K^2}{h} H + \lambda F_1 ( H^2-1  ) H \,. 
\end{align}
The above ODEs are solved numerically and shown in Fig.\,\ref{plot_HP}(c).

\begin{figure}[t!]
\centering
\mbox{
(a)
 \includegraphics[angle =-90,scale=0.3]{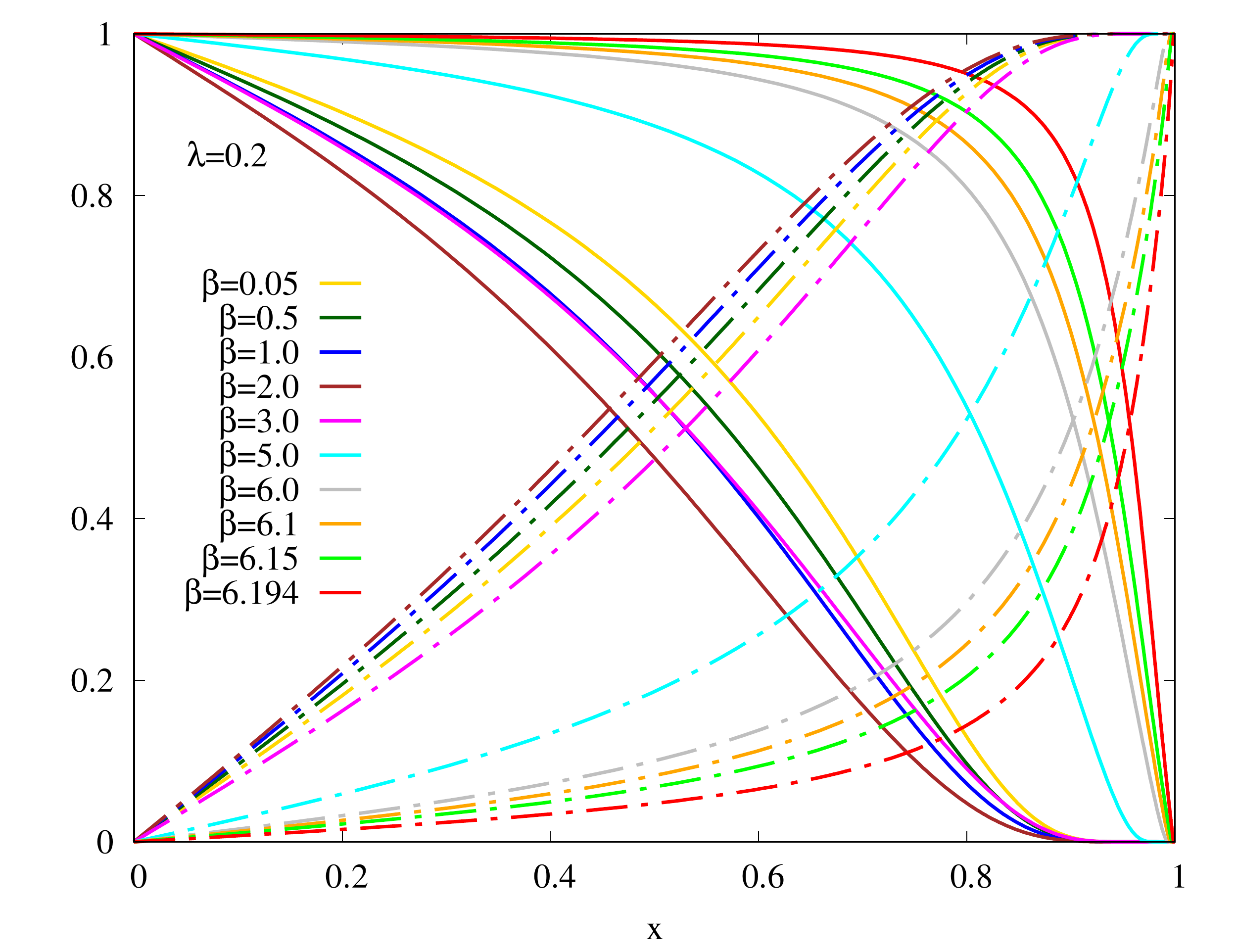}
(b)
 \includegraphics[angle =-90,scale=0.3]{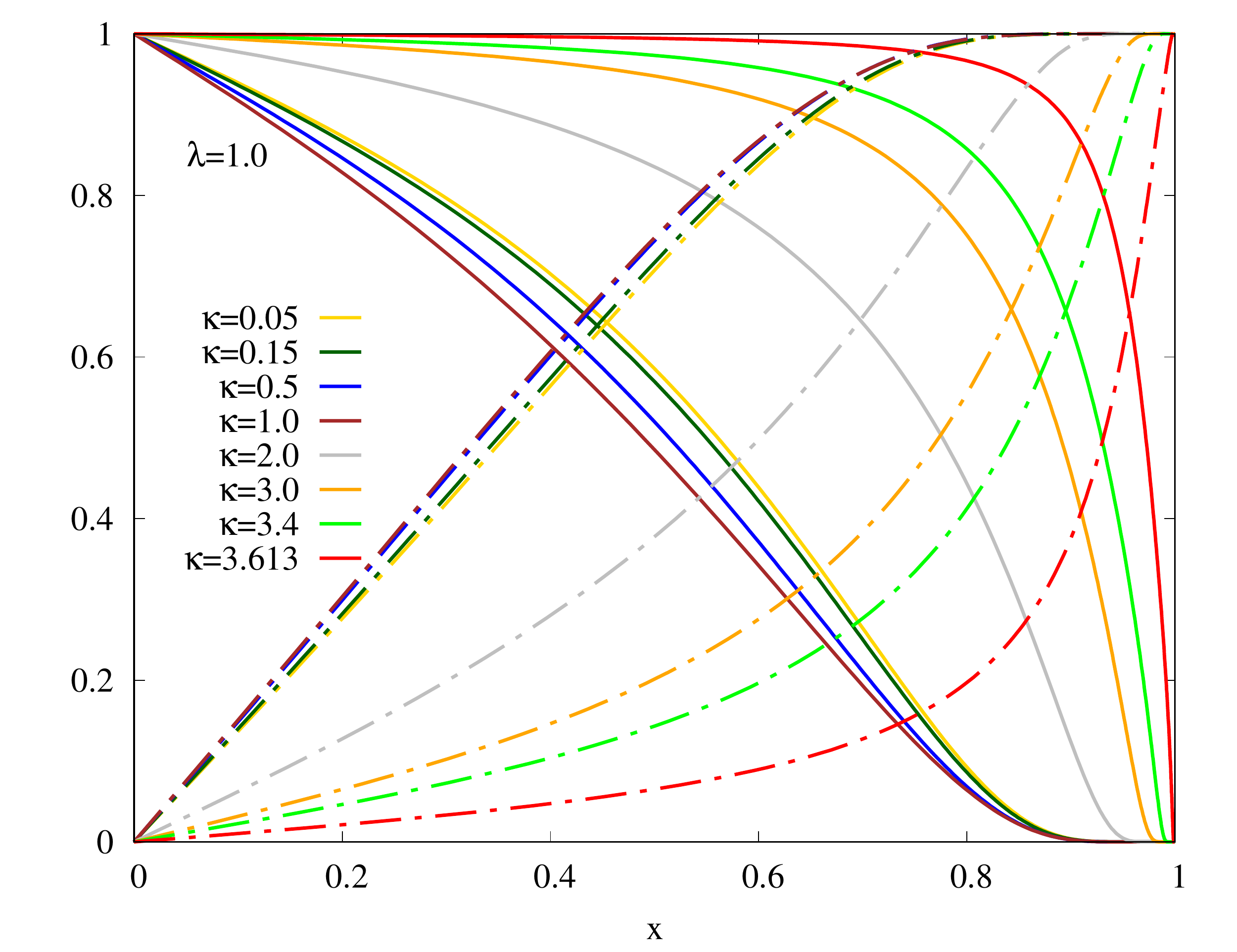}
 }
\mbox{
(c)
 \includegraphics[angle =-90,scale=0.3]{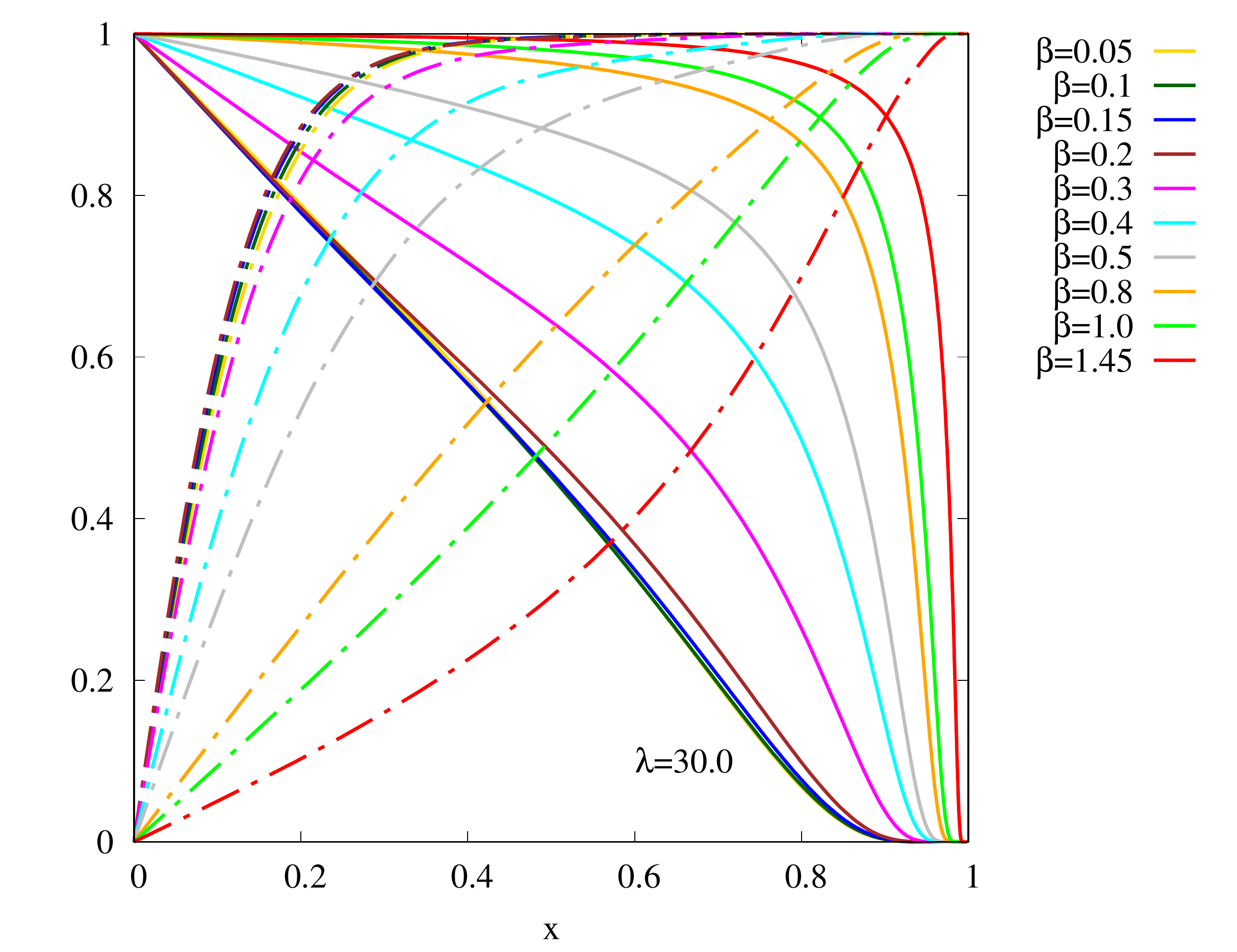}
(d)
\includegraphics[angle =-90,scale=0.3]{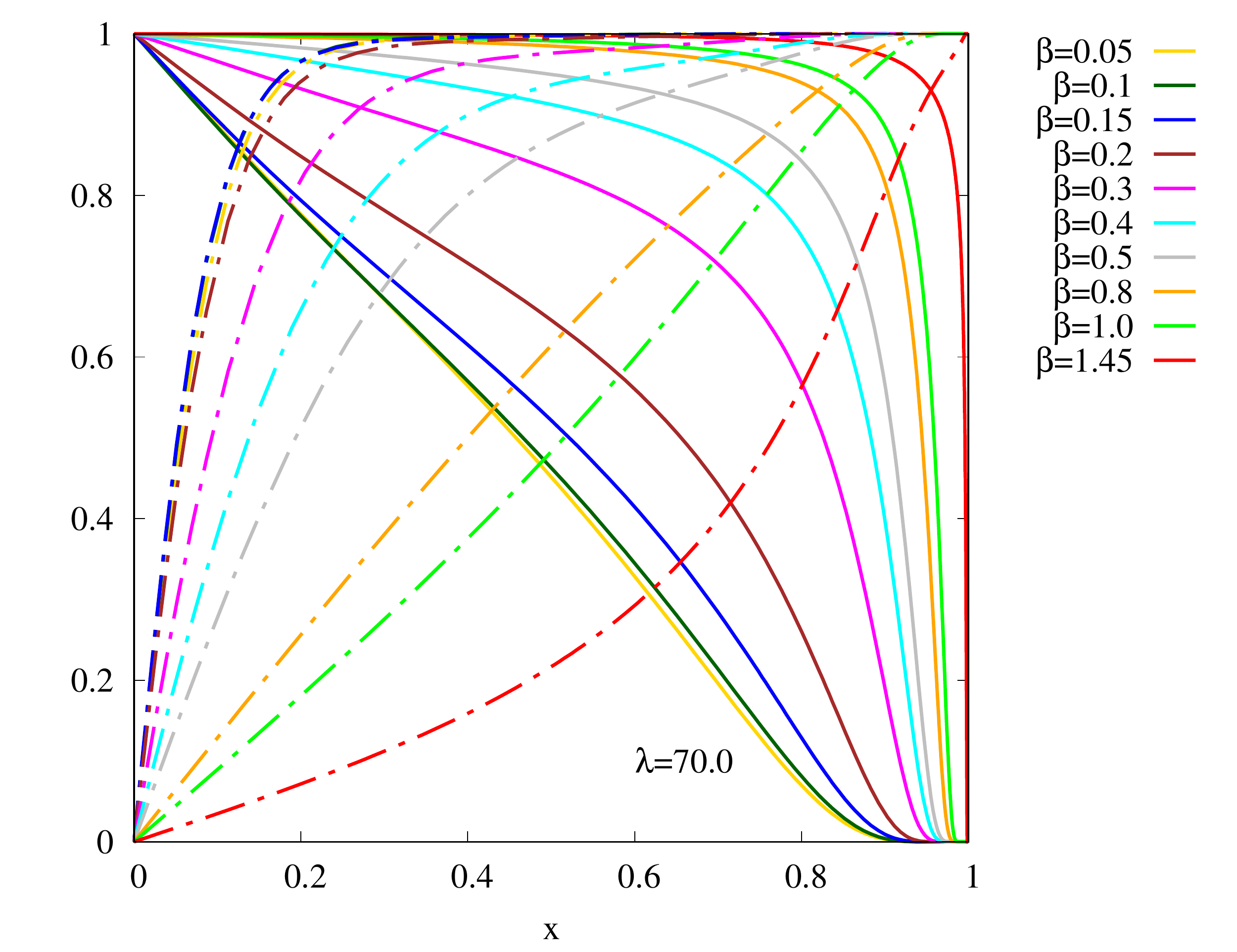}
}
\caption{The gauge field $K(x)$ (solid line) and Higgs field $H(x)$ (dash-dotted line) in the compactified coordinate $x$ for the wormhole solutions by varying $\beta$ with fixed $\lambda$: (a) $\lambda=0.2$; (b) $\lambda=1$; (c) $\lambda=30$ and (d) $\lambda=70$.}
\label{plot_wormsol_KH}
\end{figure}

\begin{figure}[t!]
\centering
\mbox{
(a)
 \includegraphics[angle =-90,scale=0.3]{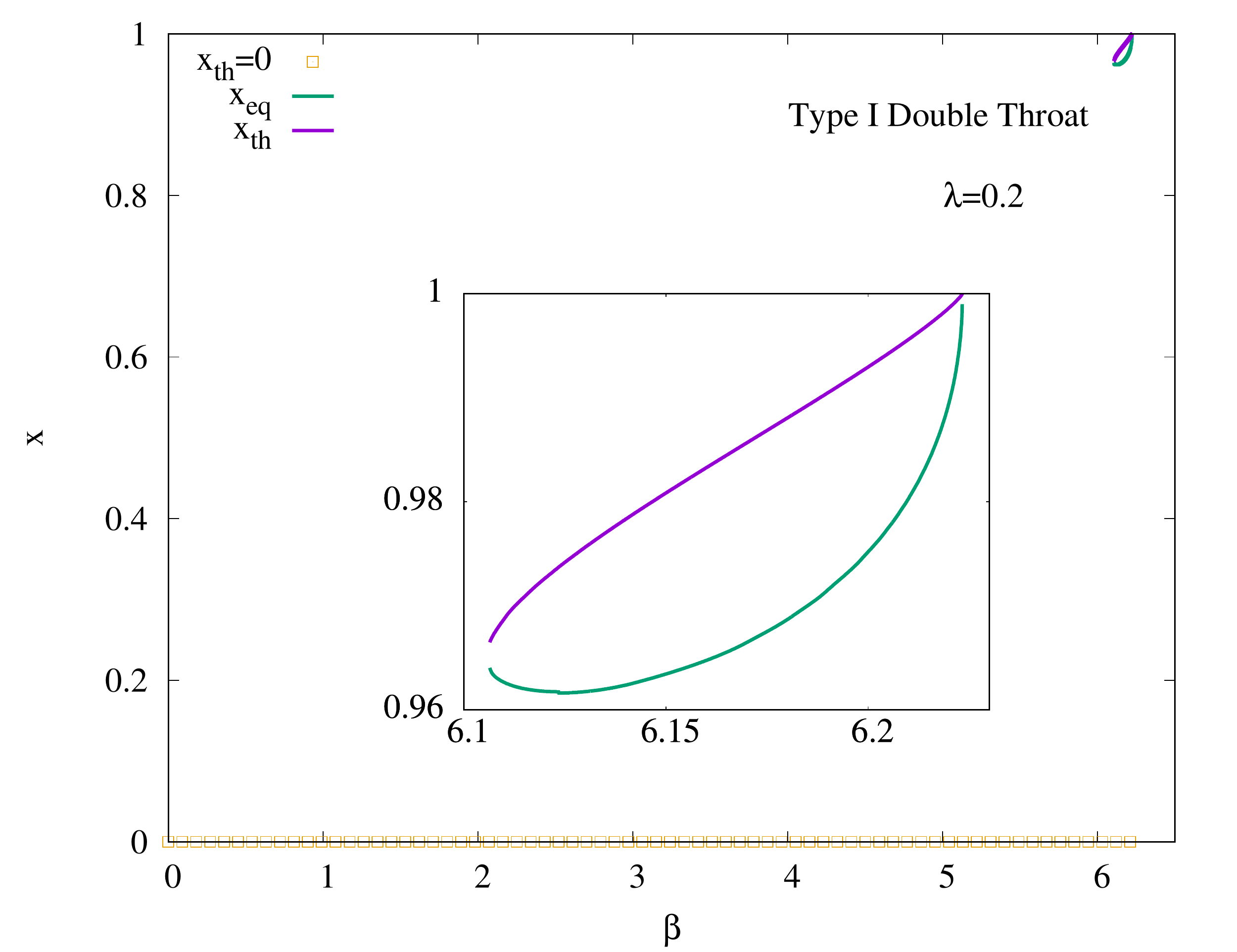}
(b)
 \includegraphics[angle =-90,scale=0.3]{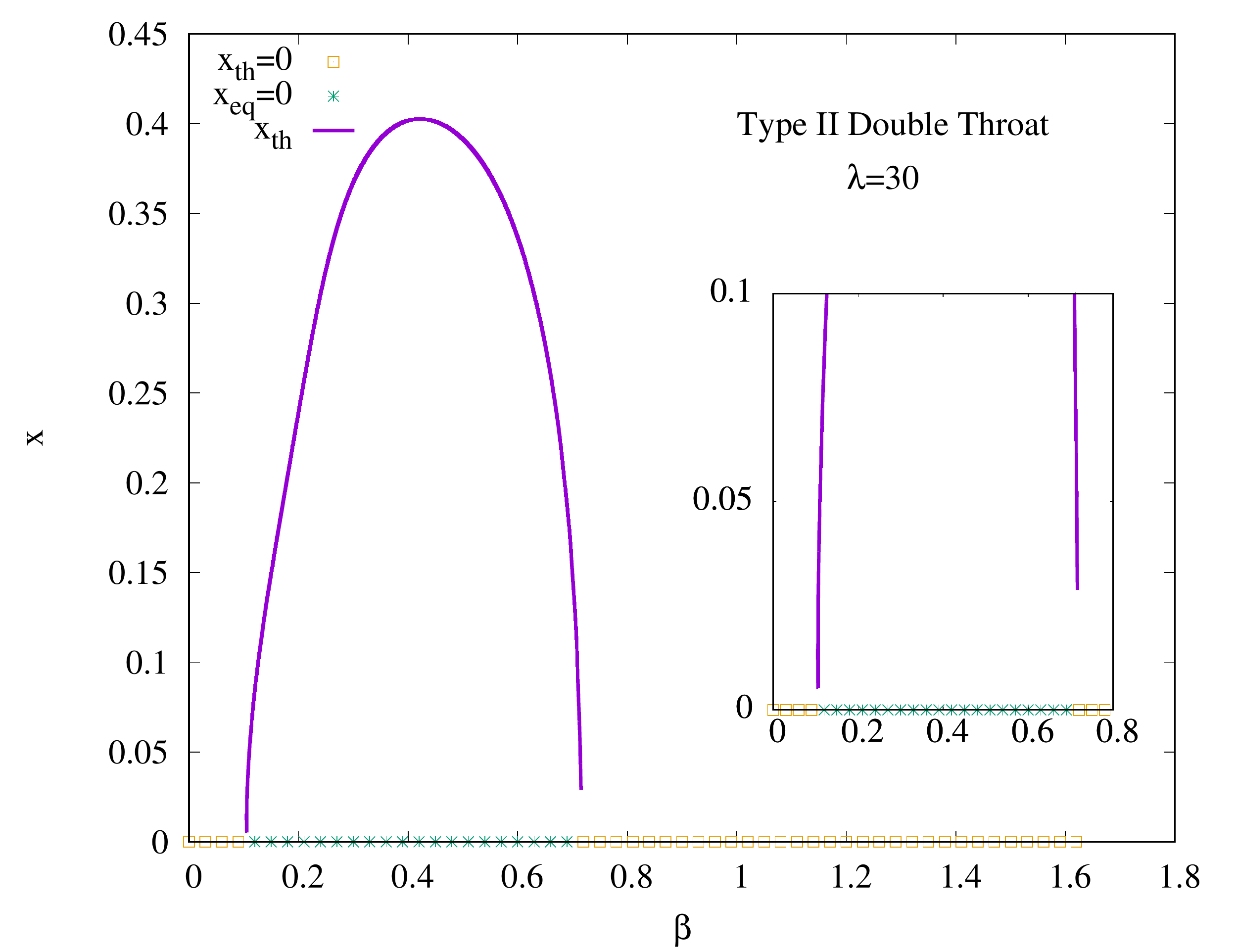}
 }
\mbox{
(c)
 \includegraphics[angle =-90,scale=0.3]{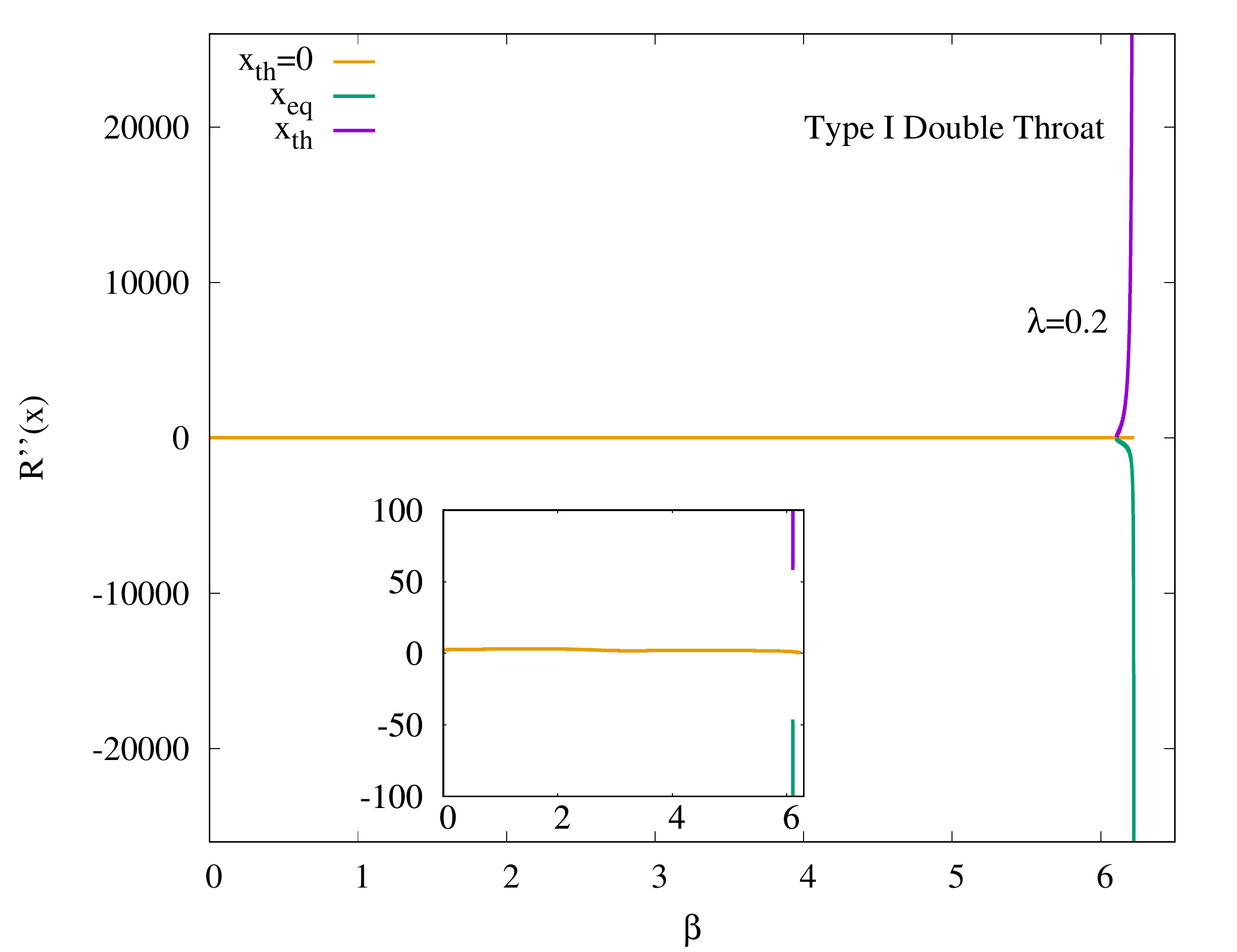}
(d)
 \includegraphics[angle =-90,scale=0.3]{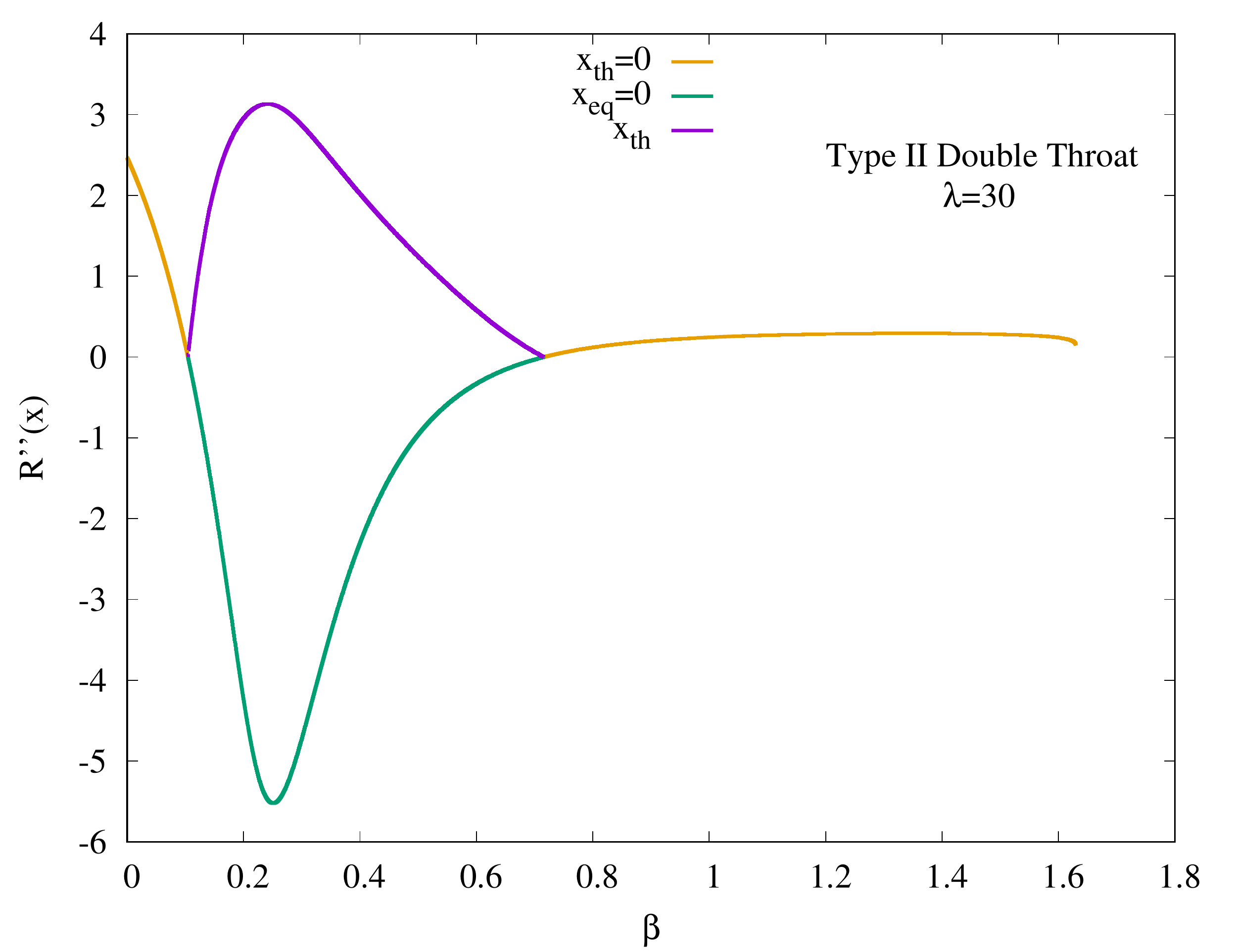}
}
\mbox{
(e)
 \includegraphics[angle =-90,scale=0.3]{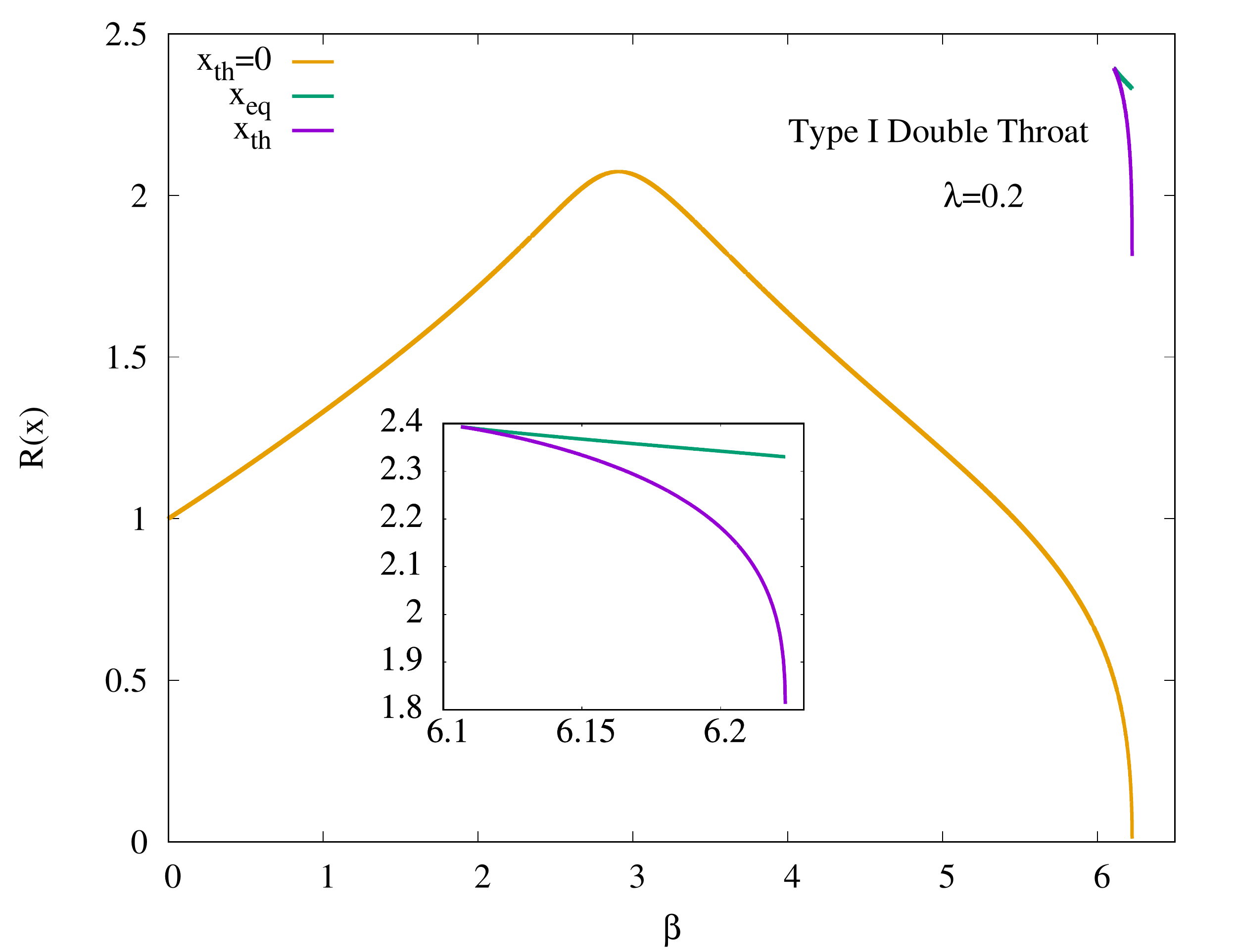}
(f)
 \includegraphics[angle =-90,scale=0.3]{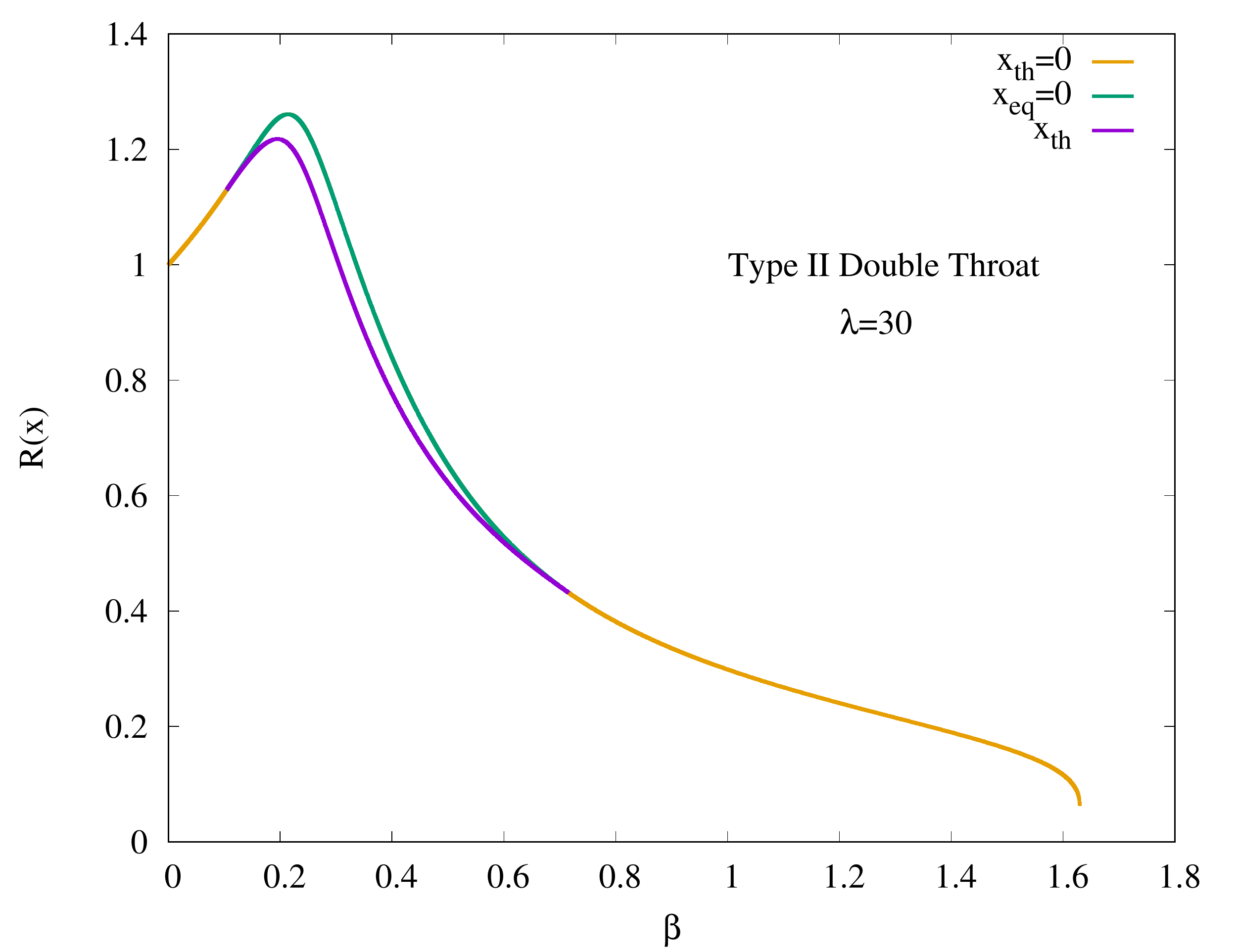}
}
\mbox{
(g)
\includegraphics[angle =-90,scale=0.3]{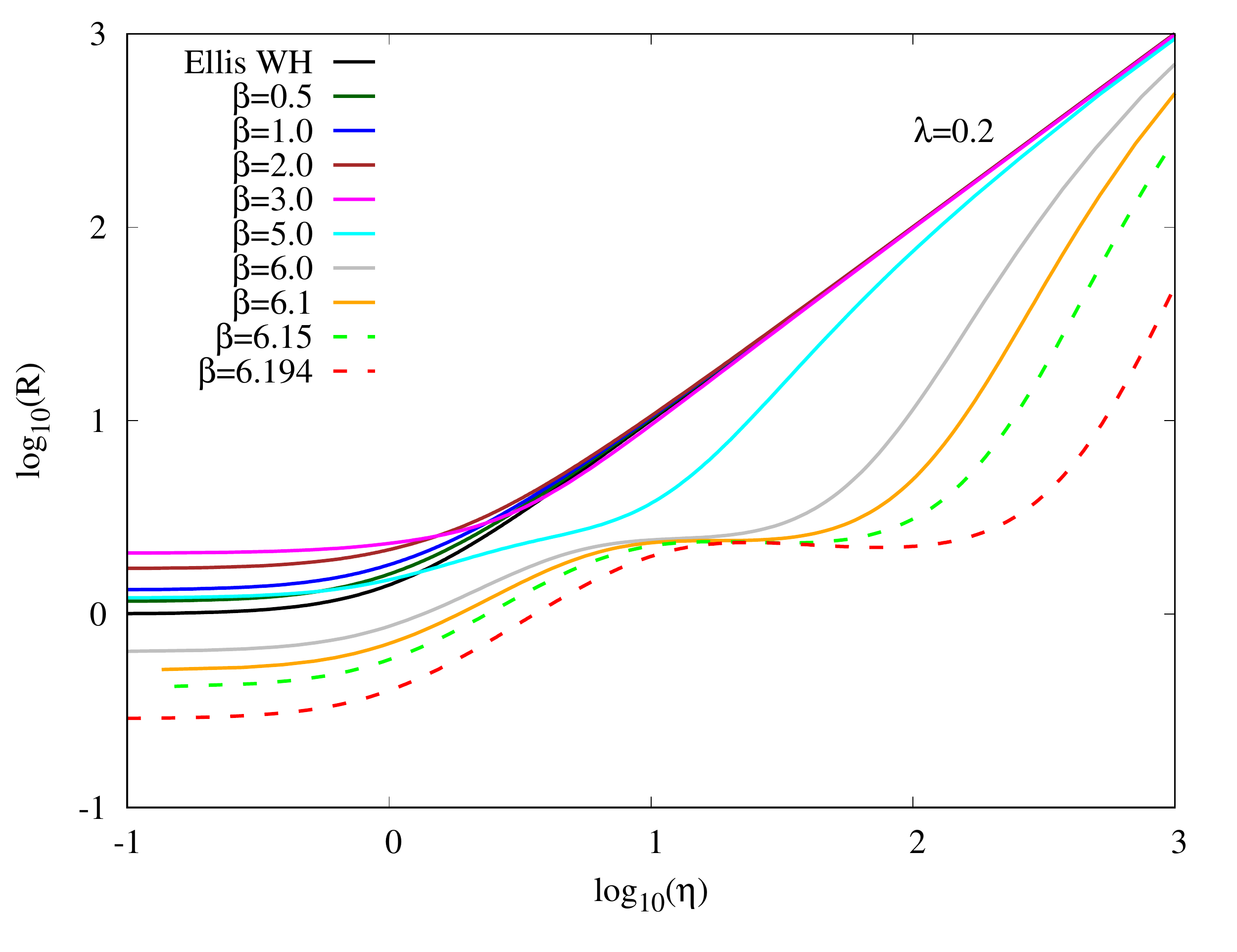}
(h)
\includegraphics[angle =-90,scale=0.3]{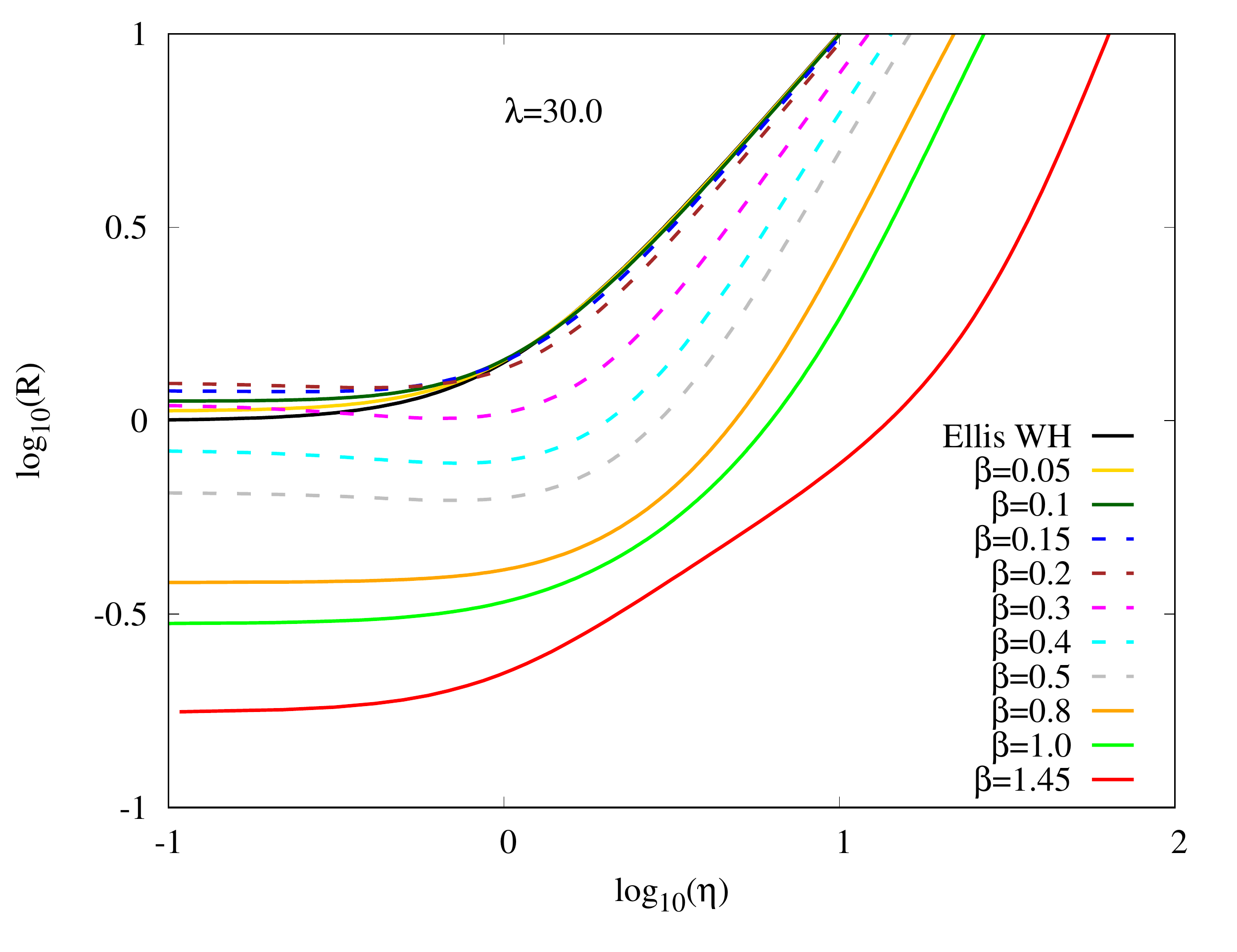}
}
\caption{The geometrical properties of the wormhole solutions. (a) and (b) are the location of throats $x_{\text{th}}$ and equator $x_{\text{eq}}$ in the compactified coordinate $x$ versus $\beta$, respectively for $\lambda=0.2$ (Type I), $\lambda=30$ (Type II); (c) and (d) are the second derivative of the circumferential radius of throats and equator versus $\beta$, respectively for $\lambda=0.2, 30$; (e) and (f) are the size of throats and equator versus $\beta$, respectively for $\lambda=0.2, 30$; (g) and (h) are the circumferential radius $R$ in the radial coordinate $\eta$ with several values of $\beta$, respectively for $\lambda=0.2, 30$; }
\label{plot_geom_throat}
\end{figure}

\begin{figure}[t!]
\centering
\mbox{
(a)
\includegraphics[angle =0,scale=0.2]{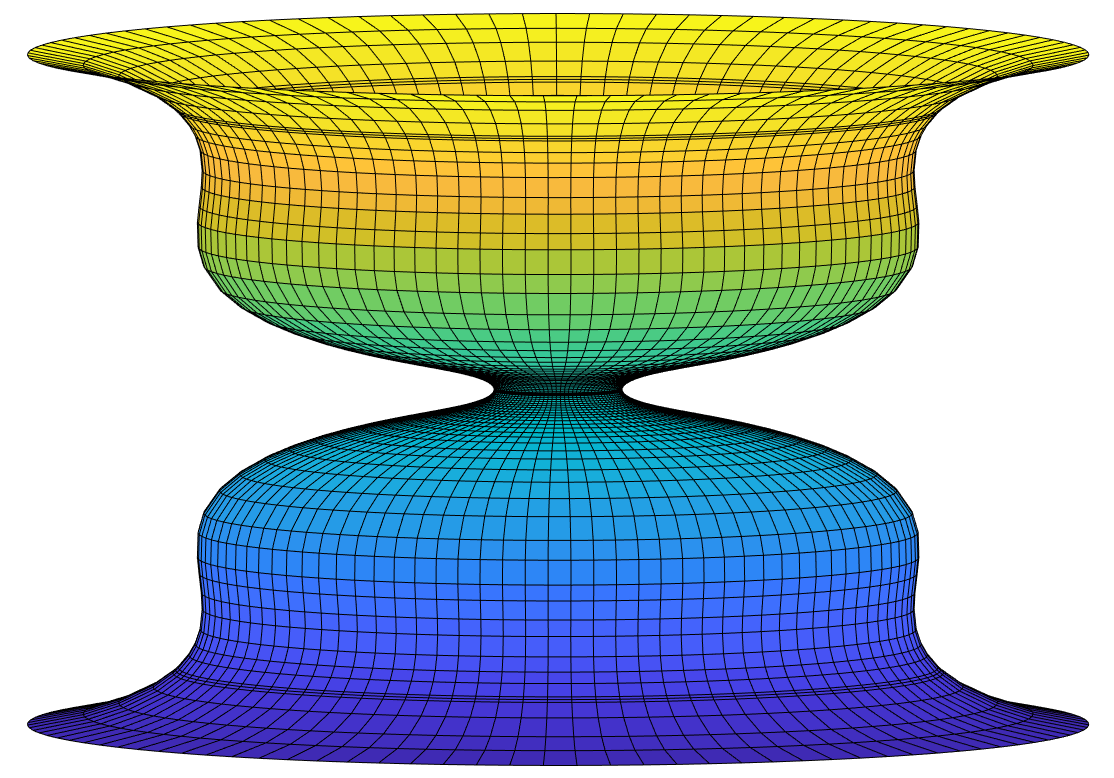} \quad \quad 
(b)
\includegraphics[angle =0,scale=0.2]{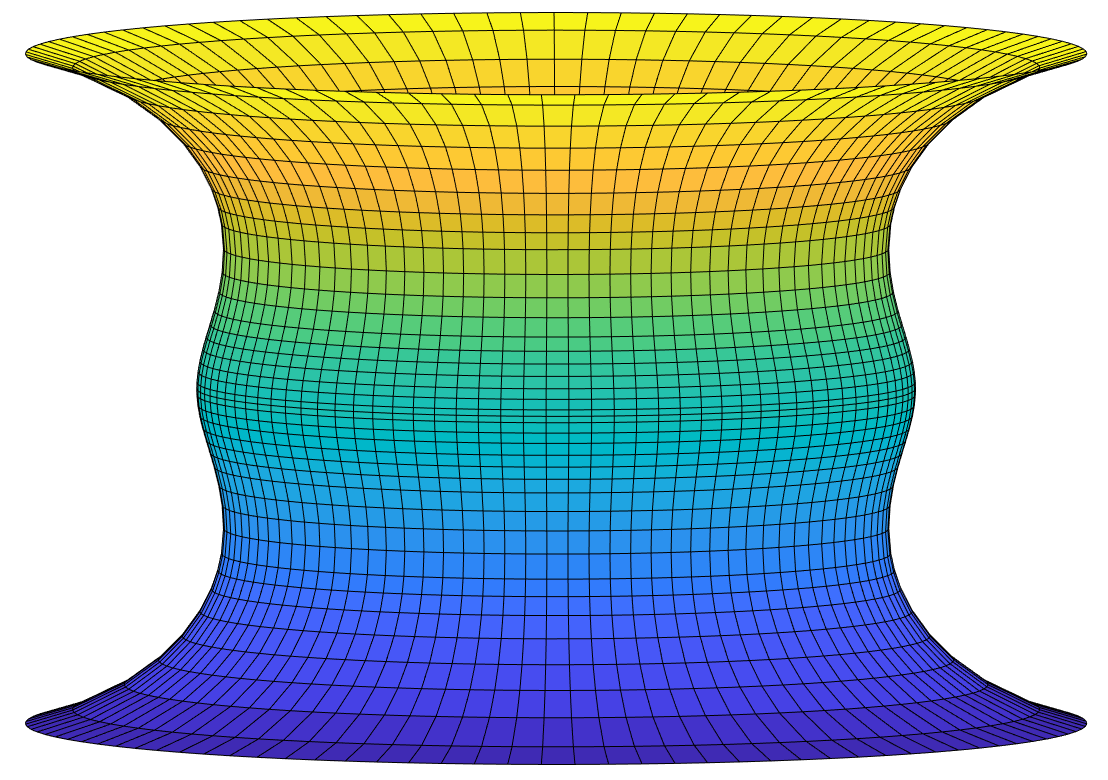}
}
\caption{The isometric embedding for the wormholes solutions: (a) Type I double throat $\lambda=0.2$ and $\beta=6.15$; (b) Type II double throat $\lambda=30$ and $\beta=0.4$.}
\label{plot_embed3d}
\end{figure}

\begin{figure}[t!]
\centering
\mbox{
(a)
\includegraphics[angle =-90,scale=0.3]{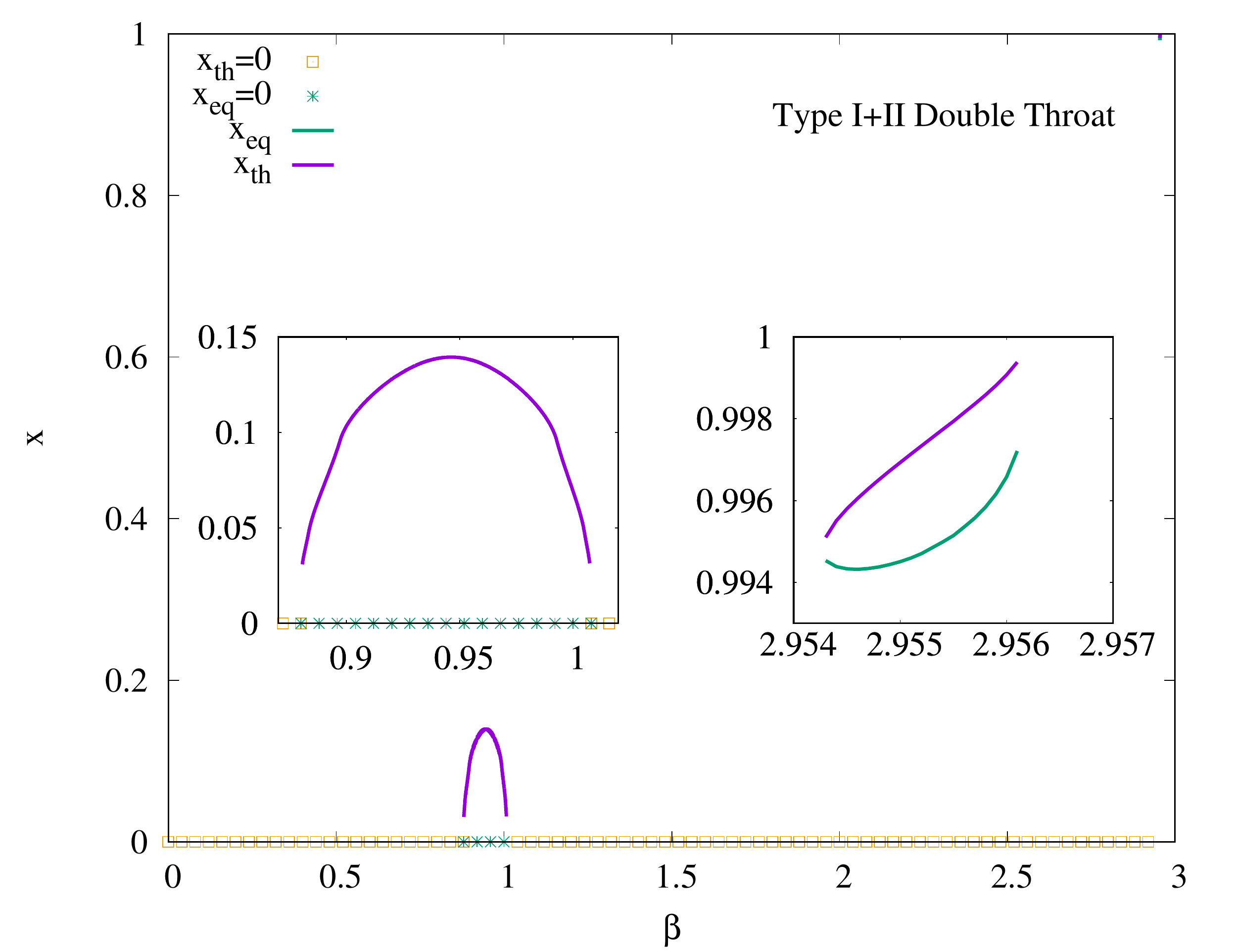}
(b)
\includegraphics[angle =-90,scale=0.3]{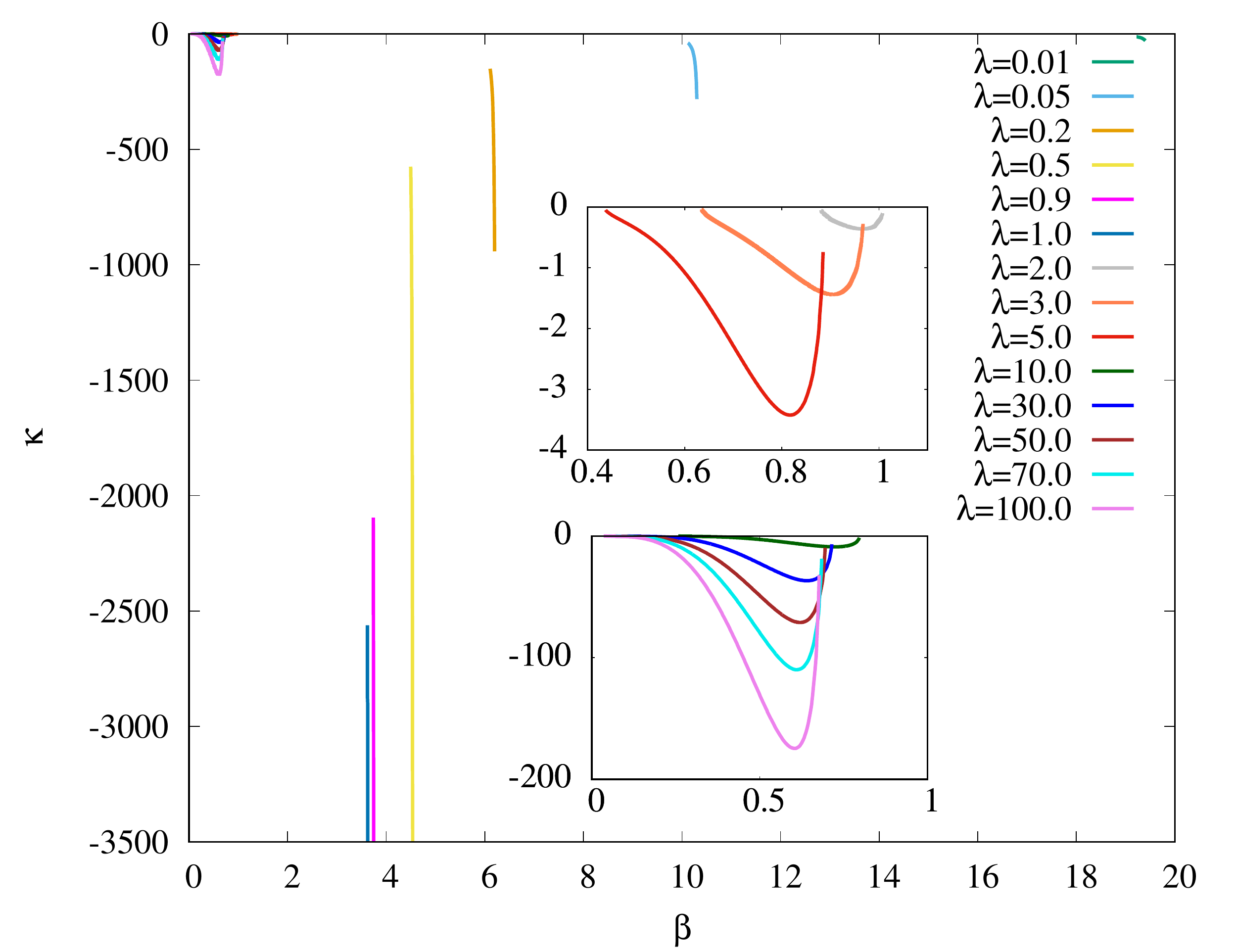}
}
\caption{(a) The location of throats and equator of hybrid type I + II double throat configuration for $\lambda=2$. (b) The surface gravity $\kappa$ at the throat of wormholes for several values of $\lambda$.}
\label{plot_geom_throat_1_2}
\end{figure}

\begin{figure}[t!]
\centering
\mbox{
(a)
 \includegraphics[angle =-90,scale=0.3]{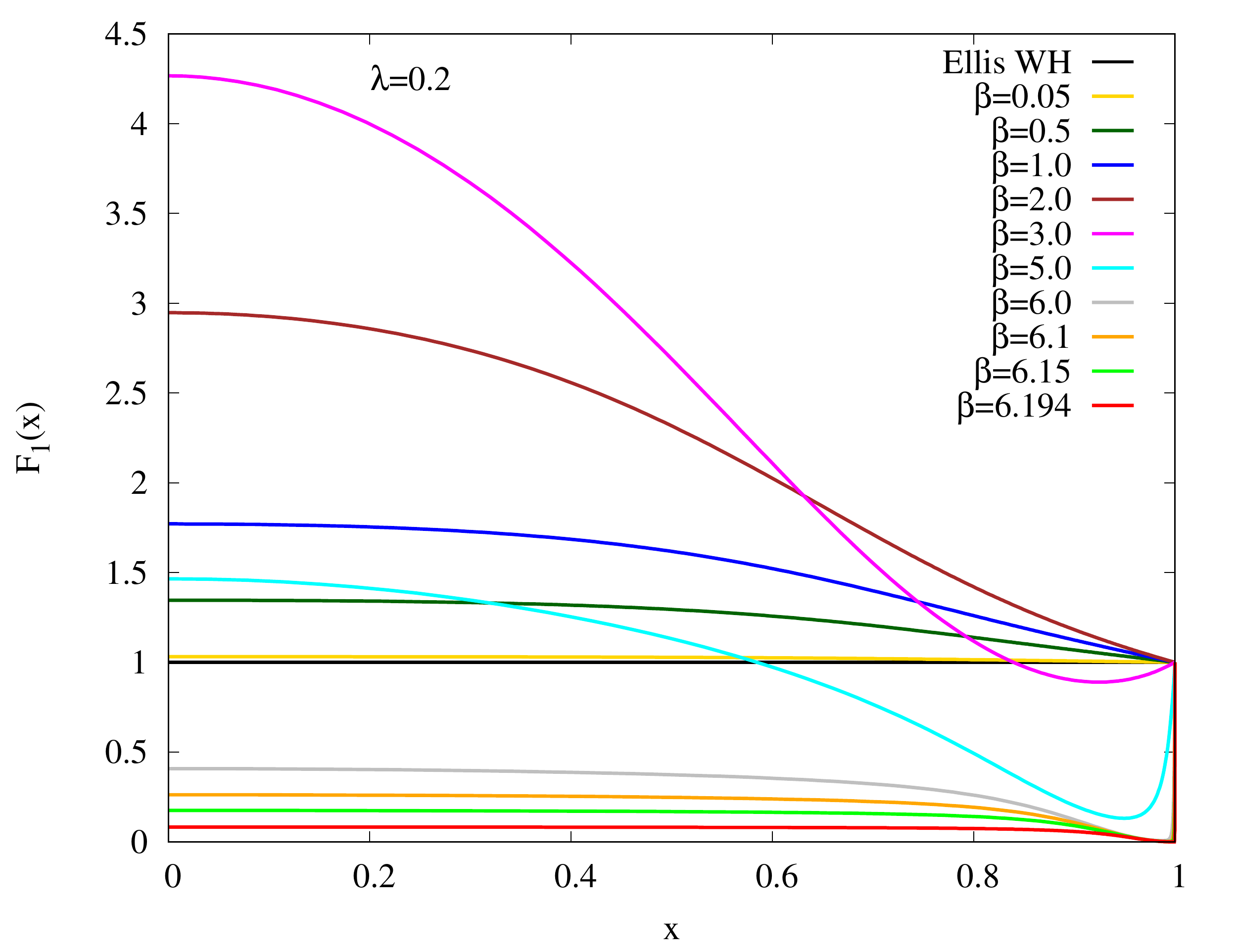}
(b)
 \includegraphics[angle =-90,scale=0.3]{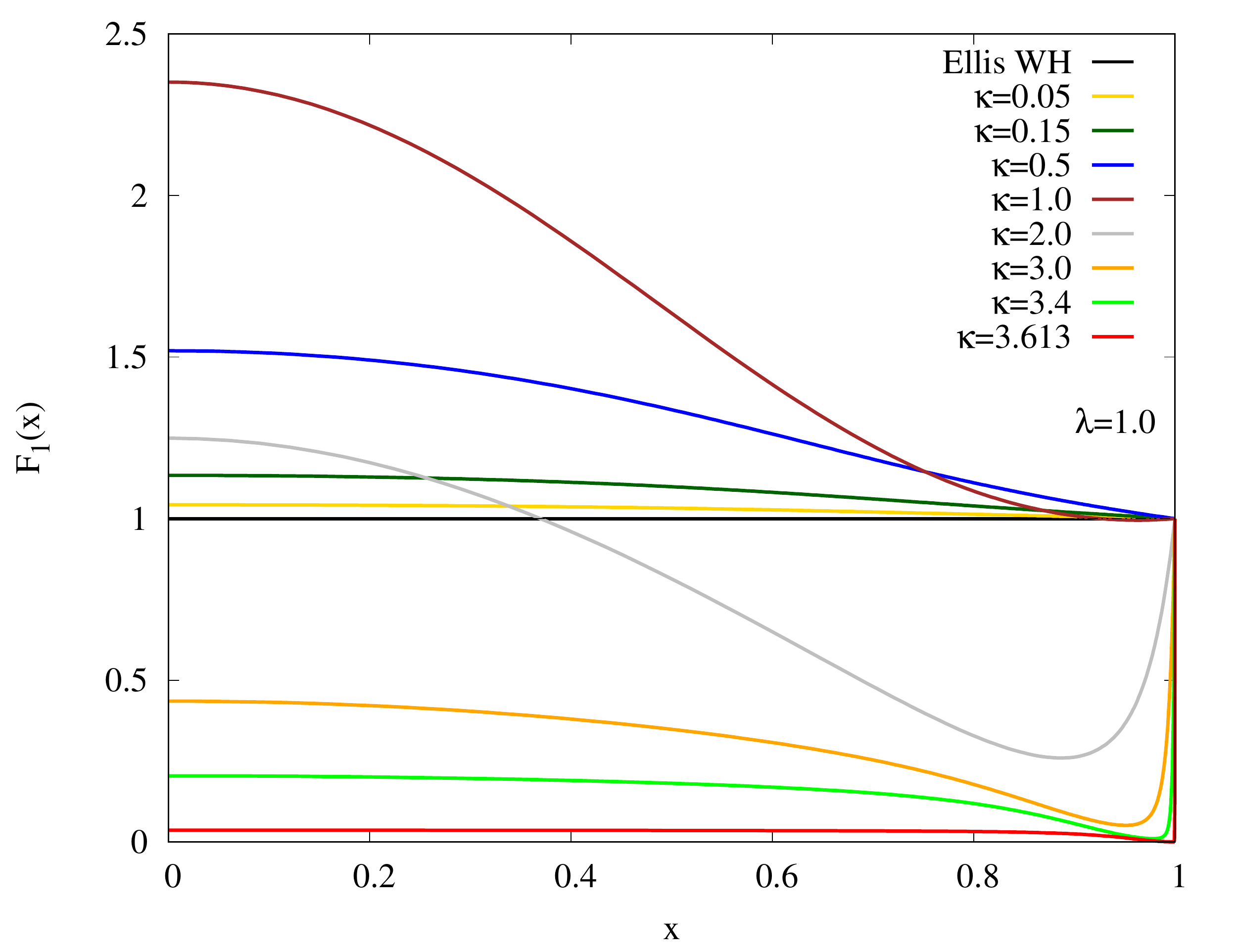}
 }
\mbox{
(c)
 \includegraphics[angle =-90,scale=0.3]{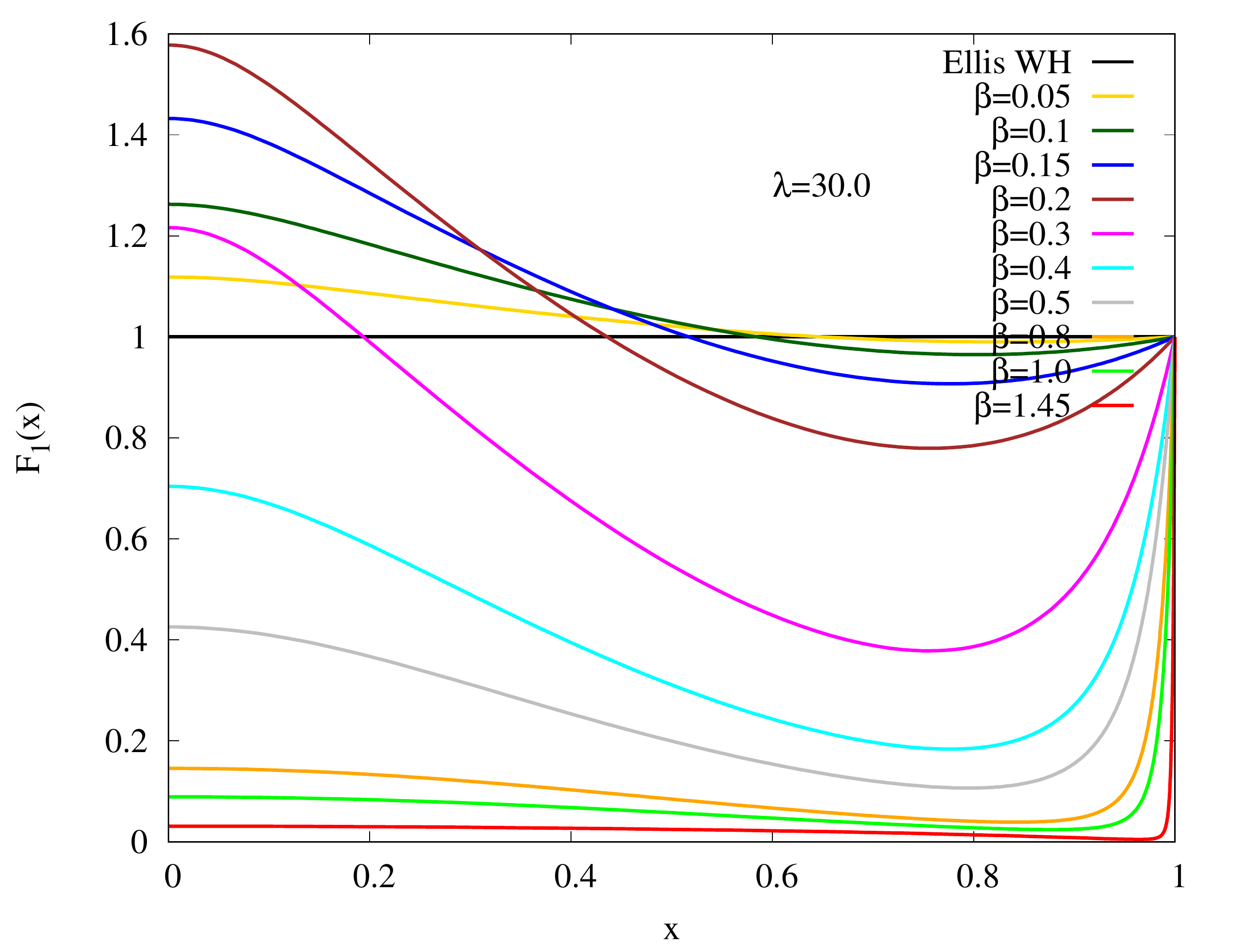}
(d)
\includegraphics[angle =-90,scale=0.3]{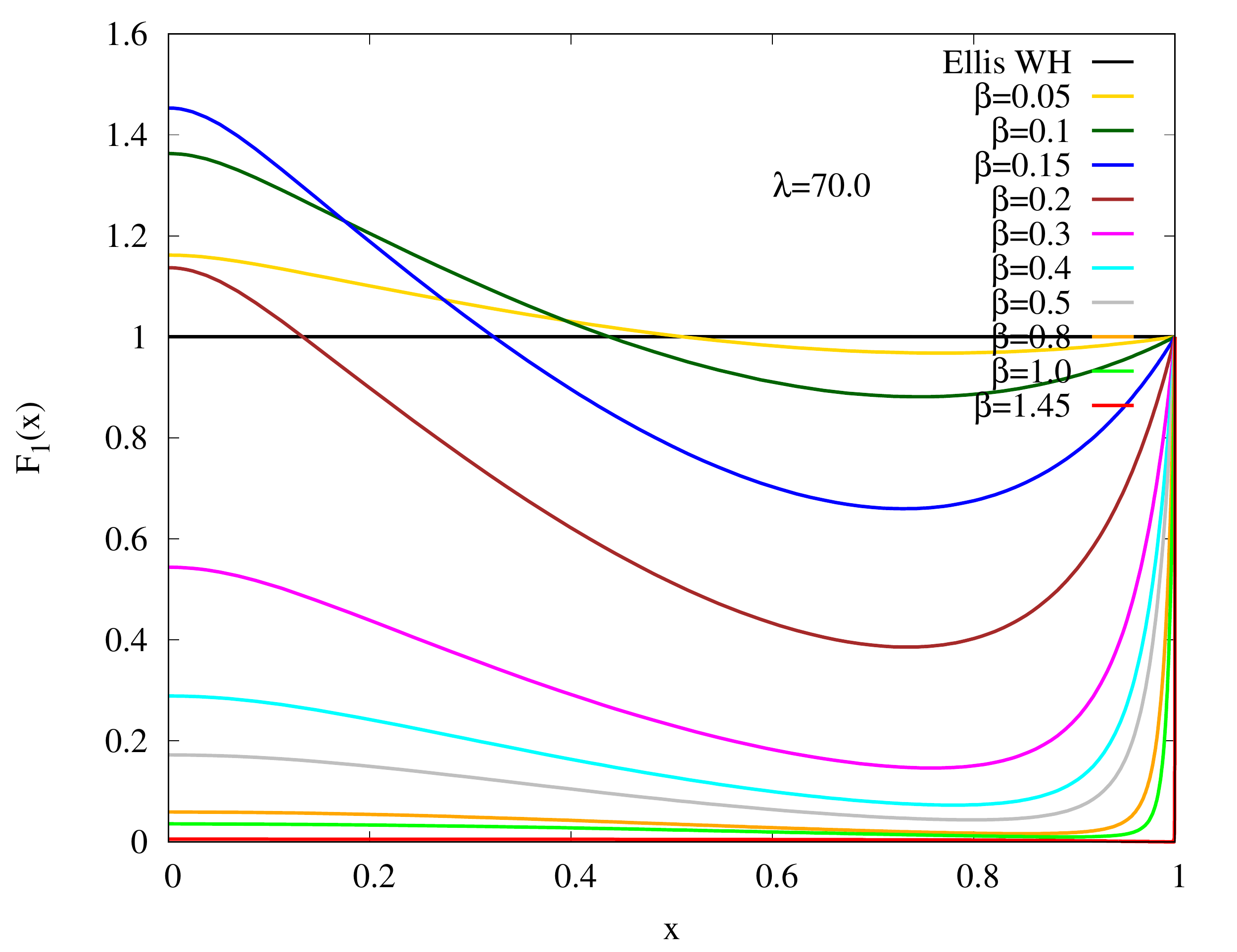}
}
\caption{The metric function $F_1(x)$ in the compactified coordinate $x$ for the wormhole solutions by varying $\beta$ with fixed $\lambda$: (a) $\lambda=0.2$; (b) $\lambda=1$; (c) $\lambda=30$ and (d) $\lambda=70$.}
\label{plot_wormsol_exf1}
\end{figure}

\begin{figure}[t!]
\centering
\mbox{
(a)
 \includegraphics[angle =-90,scale=0.3]{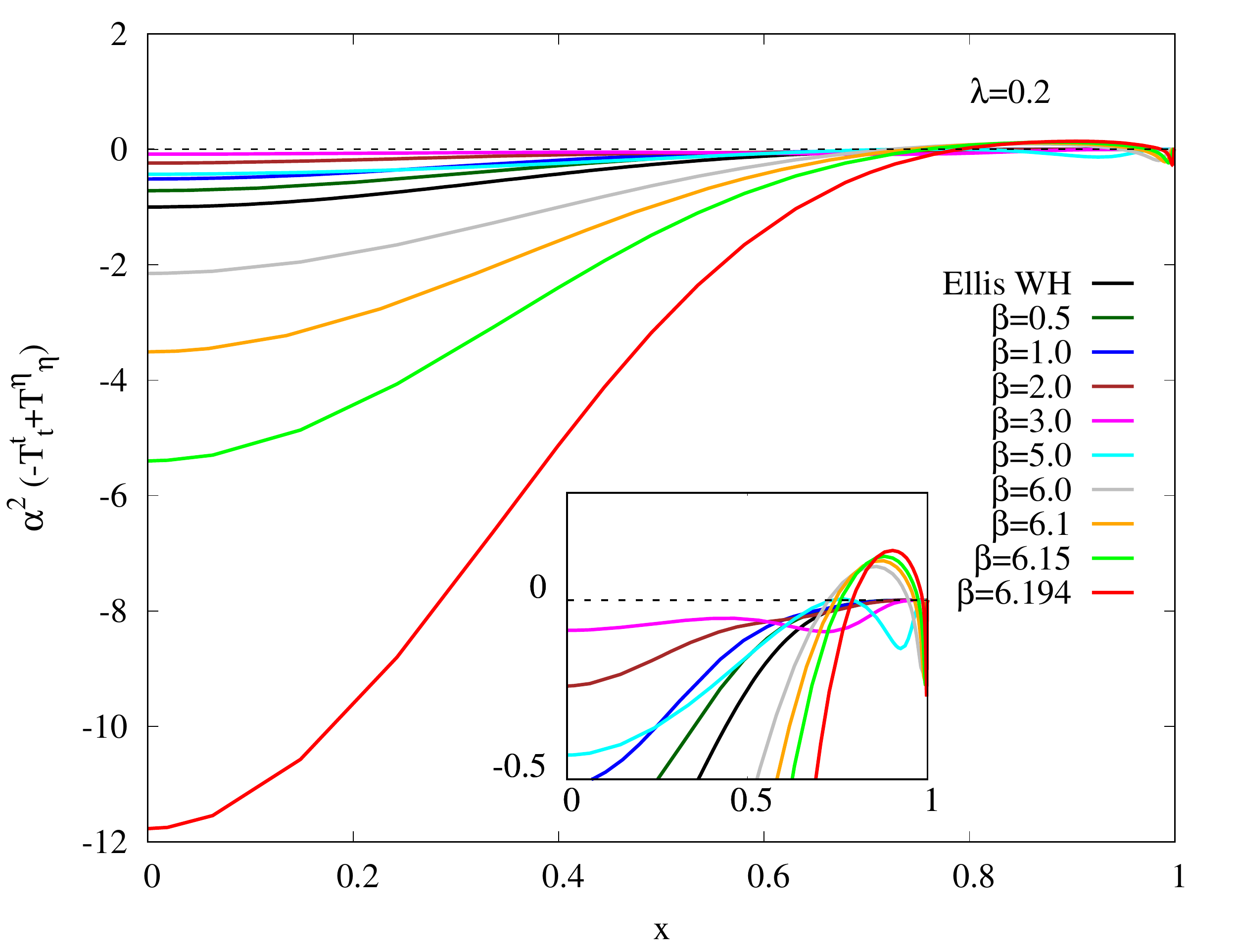}
(b)
 \includegraphics[angle =-90,scale=0.3]{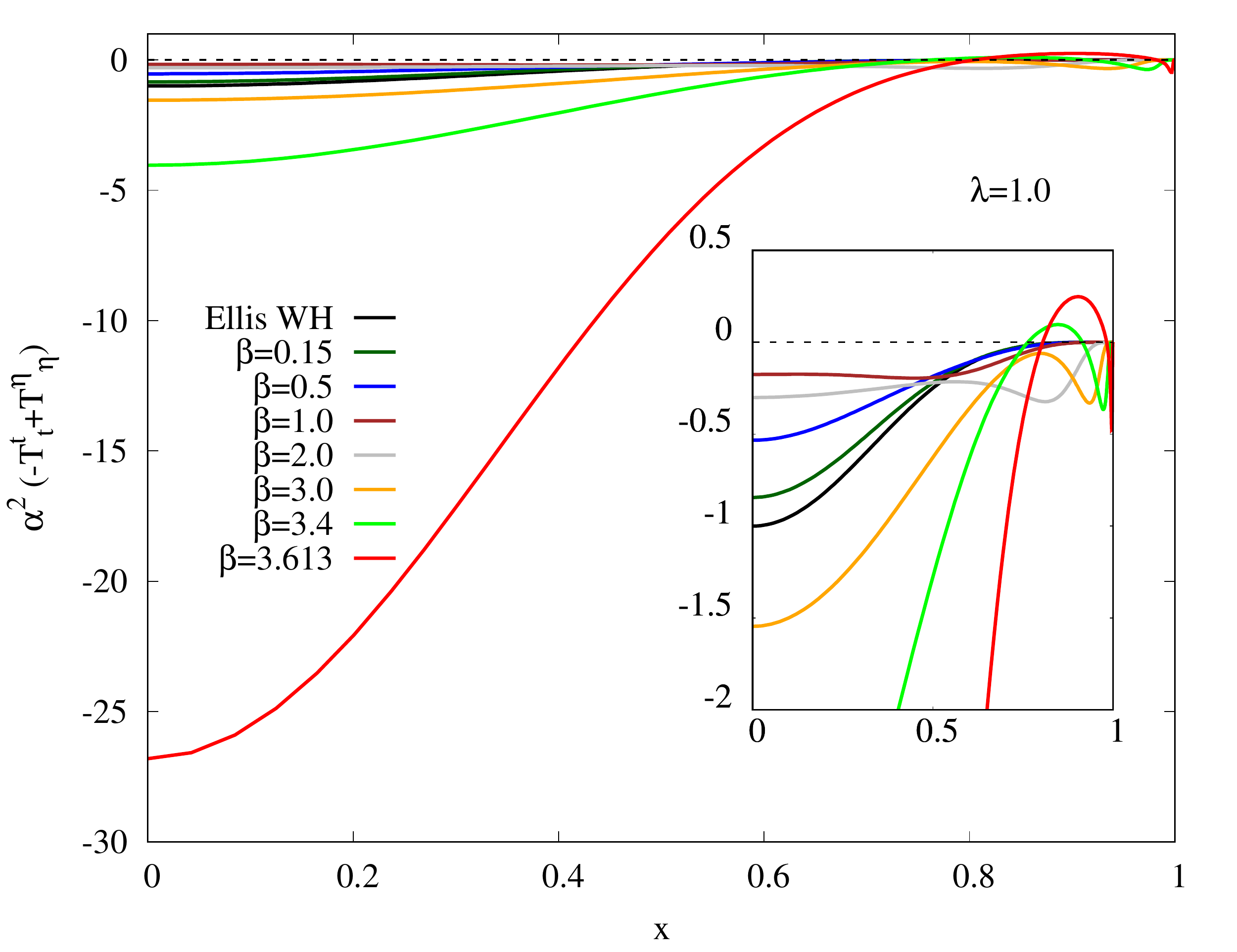}
 }
\mbox{
(c)
 \includegraphics[angle =-90,scale=0.3]{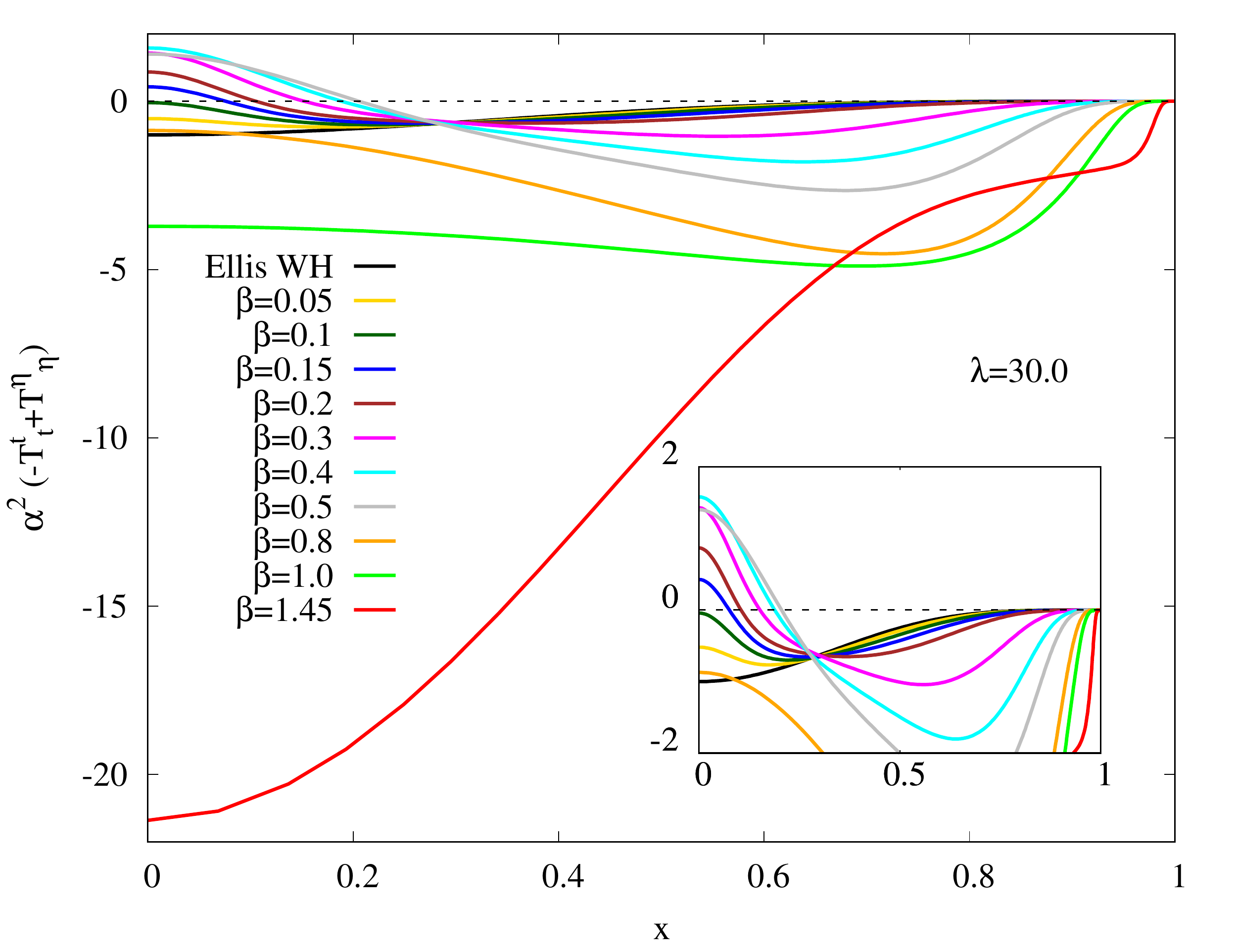}
(d)
\includegraphics[angle =-90,scale=0.3]{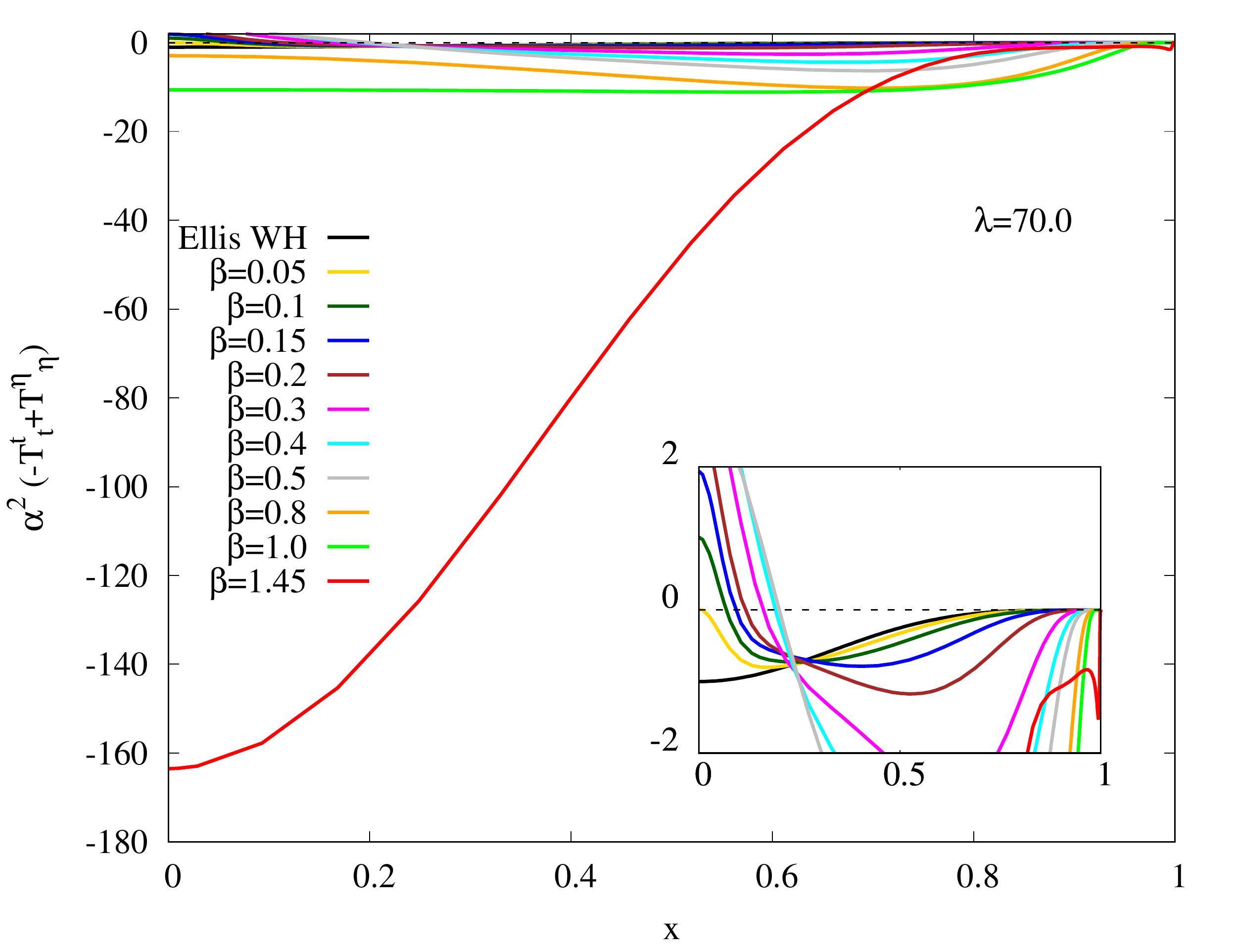}
}
\mbox{
(e)
 \includegraphics[angle =-90,scale=0.3]{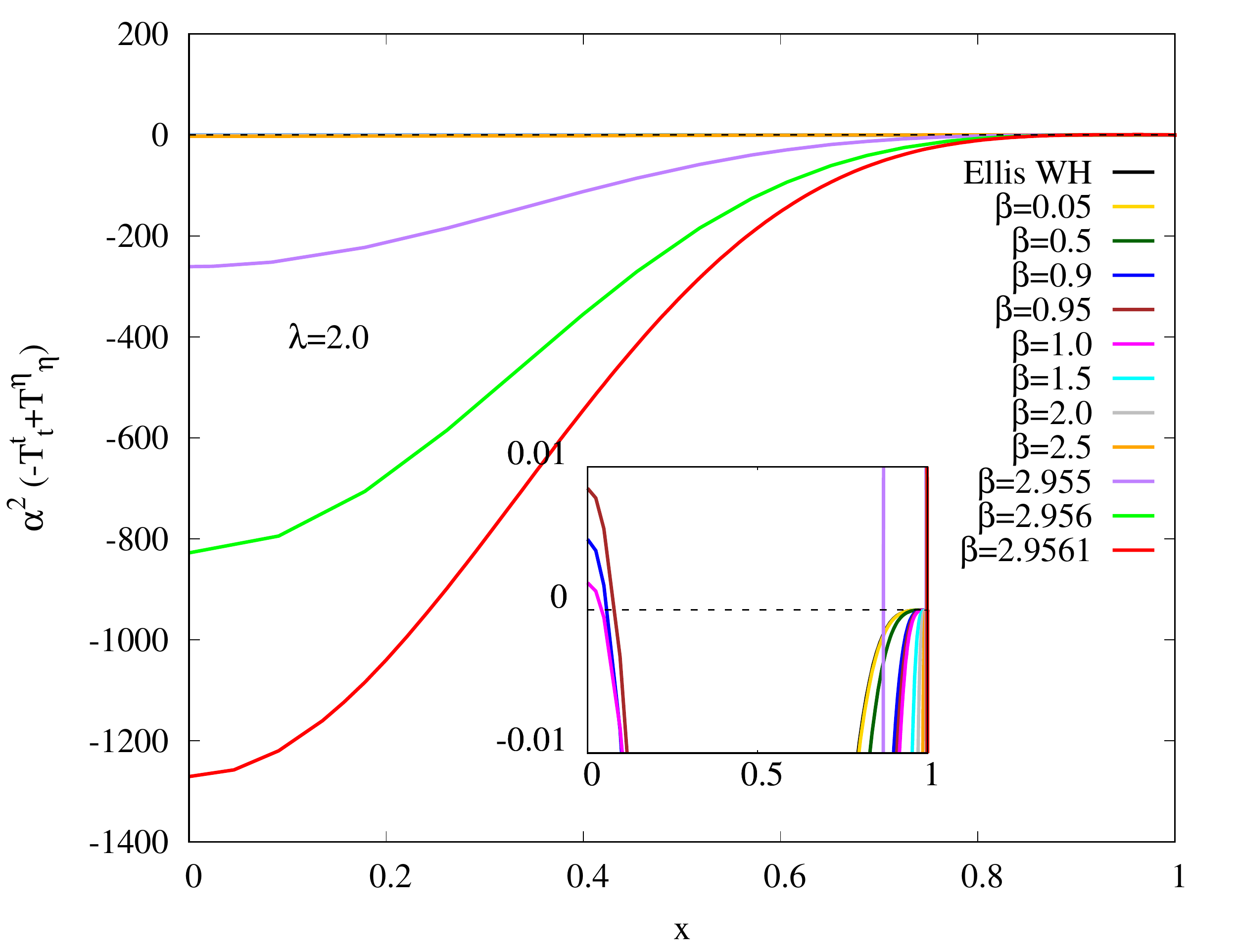}
(f)
 \includegraphics[angle =-90,scale=0.3]{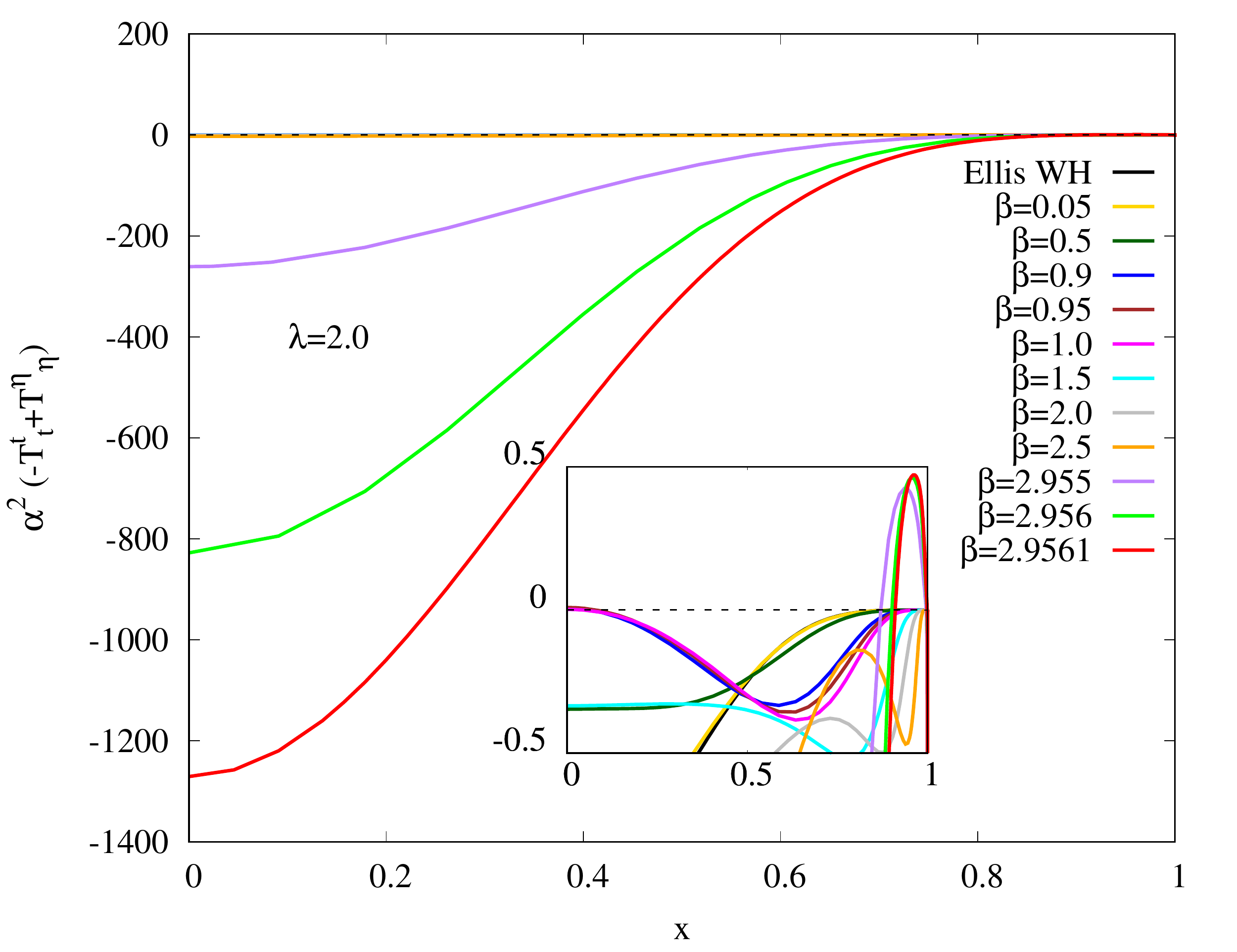}
}

\caption{The violation of null energy condition in the compactified coordinate $x$ for the wormhole solutions by varying $\beta$ with fixed $\lambda$: (a) $\lambda=0.2$; (b) $\lambda=1$; (c) $\lambda=30$; (d) $\lambda=70$; (e) $\lambda=2$ with the inset of zoom in for small postive values of NEC at $x=0$; (f) $\lambda=2$ with the inset of zoom in for the peak near $x=1$.}
\label{plot_wormsol_NEC}
\end{figure}

\subsection{With Backreaction}

Here we exhibit our numerical results for hairy wormholes beyond the BPS limit ($\lambda \neq 0$) by fixing the Higgs self-interaction parameter $\lambda$ for the range $[0,100]$ while varying the gravitational coupling constant $\beta$. The hairy wormholes solutions should exist for all positive real values of $\lambda$. The hairy wormholes in the BPS limit could take any real positive values of $\beta$ \cite{Chew:2020svi}, this is in contrast to the YM wormhole where the limiting configuration is the extremal Reissner-Nordstrom black hole for higher nodes \cite{hauser2014hairy}. Recall that when the gauge fields don't present in the probe limit, the Ellis wormhole has the analogue of Schwarzschild solution for the black holes \cite{hauser2014hairy,Chew:2020svi}. Recall also that the Ellis wormhole is massless, the circumferential radius of throat and scalar charge are unity. With fixed $\lambda$, when we increase $\beta$ from zero, the families of solutions for hairy wormholes emerge from the Ellis wormhole, thus the properties of these hairy wormholes behave differently with the Ellis wormholes when $\beta$ increases. 

Fig.\,\ref{plot_prop}(a) shows that the hairy wormholes gain the mass when scaled gravitational strength $\alpha$ ($\beta=2\alpha^2$) increases from zero for several fixed values of $\lambda$. In the BPS limit, the mass of hairy wormholes increase monotonically as $\alpha$ increases and the mass is always positive. For a small value of the Higgs self-interaction $\lambda$, the mass of wormholes increases to a maximum value when $\alpha$ increases from zero. However, their mass subsequently drops very sharply to the negative value as $\alpha$ approaches a critical value. For a large value of $\lambda$, the mass of wormholes decreases to the negative value monotonically from zero as $\beta$ increases from zero and then decreases very sharply when $\beta$ approaches the critical value. Thus, we see that the mass of wormholes for the non-BPS limit and BPS limit are quite different.

Fig.\,\ref{plot_prop}(b) shows that the scaled scalar charge $\alpha^2 D^2$ of the phantom field increases monotonically from unity as scaled gravitational strength $\alpha$ increases from zero for several fixed values of $\lambda$. Then $\alpha^2 D^2$ increases very sharply as $\alpha$ approaches a critical value. However, in the BPS limit $\alpha^2 D^2$ increases monotonically as $\alpha$ increases. Analogous to the mass, the scaled phantom charge of wormholes for beyond BPS limit and BPS limit are also quite different. Note that in Figs.\,\ref{plot_prop}(a) and (b) the domain of existence for scaled gravitational strength $\alpha$ shrinks for the wormholes beyond BPS limit when $\lambda$ increases, hence the global charges can change dramatically even for a smaller value of $\alpha$ for very large values of $\lambda$. Furthermore, the precision of wormhole solutions drop significantly when the gravitational strength approaches the critical value, thus we could not generate the solutions and study their properties beyond that critical value.

Figs.\,\ref{plot_wormsol_KH}(a)-(d) exhibit the profiles of gauge field $K(x)$ and Higgs field $H(x)$ in the compactified coordinate $x$ for $\lambda=0.2,1,30,70$, respectively with several values of $\beta$. The two fields behave monotonically. The Higgs field $H$ increases monotonically from zero to reach its asymptotic value $H=1$ at the infinity while the gauge field $K$ decreases monotonically from unity to reach its asymptotic value $K=0$ at the infinity. As $\beta$ approaches the critical value, $H$ increases very sharply and becomes very steep (with a very large gradient) near the asymptotically flat region, this implies that $H$ might diverge at the asymptotically flat region. However, $K$ approaches unity when $\beta$ reaches the critical value. Nevertheless in the BPS limit, the gauge field $K$ decays faster while the profile of gauge field $H$ doesn't vary too much when $\beta$ increases \cite{Chew:2020svi}.

Let's turn our discussion to the geometry of the hairy wormholes with non-vanishing $\lambda$. In the BPS limit ($\lambda=0$) when $\beta$ exceeds a value the hairy wormholes possess a double throat configuration where an equator is sandwiched by a throat at the radial coordinate $\eta=0$ and another throat which is near the asymptotic flat region \cite{Chew:2020svi}. When $\lambda > 0$, the hairy wormholes possess two different types of double throat configurations, which are type I for the range $0 < \lambda < 3.0283$ and type II for the range $1.8814 \leq \lambda \leq 100$ as shown in Fig.\,\ref{plot_geom_throat} (a) and (b), respectively. Interestingly, type I and II double throat configurations can coexist in the range of $1.8814 \leq \lambda < 3.0283$ where we will discuss this in detail in the later paragraphs.

We exhibit the wormhole solutions with $\lambda=0.2$ for type I double throat configuration as
 depicted in Figs.\,\ref{plot_geom_throat}(a), (c), (e) and (g). In Fig.\,\ref{plot_geom_throat}(a), the wormholes only have a single throat at $x=0$ represented by the yellow square on the horizontal axis when $\beta$ increases from zero. When $\beta$ approaches a critical value, the wormholes possess a double throat in which the throat at $x=0$ remains as the throat, but another throat $x_{\text{th}}$ (purple curve) simultaneously exists near the asymptotically flat region as shown 
in the inset of Fig.\,\ref{plot_geom_throat}(a). We observe that there is an equator $x_{\text{eq}}$ (green curve) in the inset of Fig.\,\ref{plot_geom_throat}(a) lies in between that two throats as shown inside the inset in Fig.\,\ref{plot_geom_throat}(a). Note that at the beginning of the formation of the double throat, $x_{\text{th}}$ and $x_{\text{eq}}$ are very close to each other, they move away from each other and move toward the spatial infinity $x=1$ as $\beta$ increases. When $\beta$ almost reaches the critical value, both $x_{\text{th}}$ and $x_{\text{eq}}$ approach each other again. In Fig.\,\ref{plot_geom_throat}(c), as $\beta$ increases from zero before approaches to the critical value, the second derivative of circumferential radius $R''(0)$ in the compactified coordinate $x$ for the throat at $x=0$ decreases from its maximum value to a very small value which is close to zero. When type I double throat appears, $R''(x_{\text{eq}})$ (green curve) is always negative and decreases very sharply for the equator $x_{\text{eq}}$ and $R''(x_{\text{th}})$ (purple curve) is always positive and increases very sharply for another throat $x_{\text{th}}$.

In Fig.\,\ref{plot_geom_throat}(e), the size of the throat $R(0)$ (yellow curve) at radial coordinate $x=0$ increases from unity to a maximum value and then decreases to nearly zero as $\beta$ increases from zero to the critical value. 
In the inset of Fig.\,\ref{plot_geom_throat}(e), the size of another throat $R(x_{\text{th}})$ (purple curve) and the size of equator $R(x_{\text{eq}})$ (green curve) are very close to each other at the beginning of the formation of the double throat because they are very close to each other. Then $R(x_{\text{th}})$ and $R(x_{\text{eq}})$ decrease and their difference become larger when $\beta$ approaches the critical value.

In Fig.\,\ref{plot_geom_throat}(g), the wormhole solutions with $\beta=0.5, 1.0, 2.0, 3.0$ clearly only possess a single throat at $x=0$. Although the solutions with $\beta=5.0, 6.0, 6.1$ still possess a single throat, their circumferential radii are being deformed for $0 \leq \log_{10}(\eta) \leq 2$ which is a process in developing the double throat configuration. Hence, the solutions with $\beta=6.15, 6.194$ (dashed line) which approach to critical value possess the double throat configuration because the circumferential radius $R$ contains two local minima and one local maximum. The isometric embedding of the type I double throat configuration for $\lambda=0.2$ can be visualized in Fig.\,\ref{plot_embed3d}(a).

For type II double throat configuration, we exhibit the wormhole solutions with $\lambda=30$ as depicted in Figs.\,\ref{plot_geom_throat}(b), (d), (f) and (h). As shown in Fig.\,\ref{plot_geom_throat}(b), the wormholes only possess a single throat at $x=0$ for small values of $\beta$, since $R''(0)>0$ (yellow curve) in Fig.\,\ref{plot_geom_throat}(d). As $\beta$ increases to a value in which $R''(0)$ decreases to zero in Fig.\,\ref{plot_geom_throat}(d), this implies that the wormholes are in a transition from a single throat configuration to a double throat configuration. Hence, $x=0$ becomes an equator (green dot in Fig.\,\ref{plot_geom_throat}(b)) with $R''(0) < 0$ (green curve) in Fig.\,\ref{plot_geom_throat}(d), and simultaneously another throat $x_{\text{th}}$ (purple curve) in Fig.\,\ref{plot_geom_throat}(b) with $R''(x_{\text{th}})>0$ (purple curve) in Fig.\,\ref{plot_geom_throat}(d) is developed nearby $x=0$. As $\beta$ continues to increase, the throat $x_{\text{th}}$ moves away from the equator $x=0$ to a maximum distance and then moves back toward the equator $x=0$ at a certain distance, while in Fig.\,\ref{plot_geom_throat}(d) $R''(0)$ (green curve) decreases from zero to a minimum value then increases again to zero and $R''(x_{\text{th}})$ increases from zero to a maximum value then decreases again to zero. Hence, both $R''(0)=R''(x_{\text{th}})=0$, then this again implies the wormholes are in another transition from a double throat configuration to a single throat configuration where in Fig.\,\ref{plot_geom_throat}(b) the throat $(x_{\text{th}})$ disappears and the equator $x=0$ changes back as a throat until $\beta$ reaches to the critical value. Moreover, we find that when $\lambda$ increases, type II double throat appears earlier for a small value of $\beta$.

In  Fig.\,\ref{plot_geom_throat}(f), the size of the throat $R(0)$ (yellow curve) at $x=0$ increases from unity as $\beta$ increases from zero. At the beginning of the formation of type II double throat configuration where the throat at $x=0$ becomes the equator with equator size $R(0)$ (green curve) and another throat $x_{\text{th}}$ with throat size $R(x_{\text{th}})$ (purple curve) simultaneously appears nearby $x=0$, hence $R(0)=R(x_{\text{th}})$. As $\beta$ continues to increase, the equator size $R(0)$ increases to a maximum value then decreases while $R(x_{\text{th}})$ also increases to a maximum value where $x_{\text{th}}$ is farthest from $x=0$ and then decreases. Note that the difference between $R(x_{\text{th}})$ between $R(0)$ is very small. When the wormholes transform back from double throat configuration to the single throat configuration, $R(0)$ and $R(x_{\text{th}})$ coincide again and then $R(0)$ decreases toward zero when $\beta$ reaches the critical value.  

In Fig.\,\ref{plot_geom_throat}(h), the wormhole solutions with $\beta=0.05, 0.1$ only possess a single throat at $x=0$ but $\beta=0.15, 0.2,0.3,0.4,0.5$ (dashed line) possess type II double throat configuration with one local minimum and one local maximum. Although the wormholes with $\beta=0.8,1.0, 1.45$ only possess a single throat, we see that their circumferential radius has been slightly deformed and eventually the circumferential radius satisfies the relation $|R| \rightarrow |\eta|$ at infinity. The isometric embedding for type II double throat configuration with $\lambda=30$ can be visualized in Fig. \ref{plot_embed3d}(b).

Previously we have mentioned that type I and II double throat configurations can coexist for $1.8814 \leq \lambda < 3.0283$, here we denote the hybrid of type I and type II as type I + II configuration and illustrate them by choosing $\lambda=2$ in Fig.\,\ref{plot_geom_throat_1_2}(a). In fact, this type I + II configuration serves as a transition state where the type I double throat configuration can transform into the type II double throat configuration gradually. Initially, the hairy wormholes only possess type I double throat configuration for $\lambda < 1.8814$ when $\beta$ approaches a critical value. When $\lambda=1.8814$, the type II double throat configuration starts to appear in addition to type I for a certain range of gravitational strength which is far lesser than that critical value. The example $\lambda=2$ has shown that the type I + II double throat configurations coexist when type II appears for $0.8806 \leq \beta \leq 1.0074$ in the left inset of Fig.\,\ref{plot_geom_throat_1_2}(a). As $\lambda$ increases from $1.8814$ to $3.0283$, the type I double throat configuration is diminishing gradually whereas the type II configuration is enlarged increasingly. We observe that in this process, the range of $\beta$ for the appearance of type I double throat configuration decreases and the locations of the new throat and equator are approaching infinity. All these factors combined contribute to the disappearance of type I double throat configuration. Eventually, when $\lambda = 3.0283$, the type I double throat configuration disappears completely and the double throat configuration is now dominated by type II. Hence, the varying of Higgs self-interaction value in the Higgs potential gives rise to the transition of type I to type II via the intermediate state of type I + II configuration.

The surface gravity $\kappa$ of wormholes is depicted in Fig.\,\ref{plot_geom_throat_1_2}(b). Recall that $\kappa$ vanishes when the wormholes only possess a single throat at $x=0$. For type I double throat configuration, $\kappa$ only assumes a finite value when $\beta$ approaches the critical value where the wormholes develop another throat $x_{\text{th}}$ near the asymptotically flat region. As $\lambda$ increases, $\kappa$ decreases very sharply. For type II double throat configuration, $\kappa$ also only assumes finite value when double throat configuration exists. Thus, starting from the beginning of the formation of double throat configuration, $\kappa$ decreases from zero to a minimum value and then increases to zero again where the wormholes transform back to their original single throat configuration. Likewise, in the BPS limit, the surface gravity of the single throat wormhole vanishes but the double throat wormhole assumes a positive finite value.

Figs.\,\ref{plot_wormsol_exf1}(a)-(d) exhibit the metric function $F_1(x)$ in the compactified coordinate $x$ for $\lambda=0.2, 1, 30, 70$, respectively. When $\beta=0$, $F_1(x)=1$ for Ellis wormholes. As $\beta$ increases, $F_1(0)$ assumes the maximum value and $F_1(x)$ decreases monotonically to its asymptotic value $F_1(1)=1$. After $\beta$ exceeds a value, $F_1$ develops another local minimum which is near the asymptotically flat region while $F_1(0)$ still retain as the local maximum. As $\beta$ increases and then approaches the critical value, both local maximum $F_1(0)$ and local minimum decrease near to zero and the function $F(x)$ increases very steeply from that local minimum toward the asymptotic value $F(1)=1$. In the BPS limit \cite{Chew:2020svi}, the $F_1$ function is strictly declining from the maximum value at $\eta=0$ to the asymptotic value.

Lastly we address the violation of scaled null energy condition (NEC) in the compactified coordinate $x$ demonstrated in Figs.\,\ref{plot_wormsol_NEC}(a)-(e) for $\lambda=0.2, 1, 30, 70, 2$, respectively. The closed form of the expression for the NEC can be found in our previous work \cite{Chew:2020svi}. For all cases of $\lambda$, the violation of scaled NEC at the throat of wormholes $x=0$ initially is maximum when $\beta=0$ where the corresponding wormhole is Ellis wormhole (black curve). As $\beta$ increases, the violation of scaled NEC decreases but then increases again and the scaled NEC becomes most violated when $\beta$ approaches the critical value at the end. For $\lambda=0.2, 1$, the NEC at $x=0$ can be minimally violated because it reaches a very small number which is close to zero.   
Besides, we observe that the appearance of double throat configuration can affect the violation of NEC. For $\lambda=0.2, 1$, when $\beta$ approaches the critical value, both cases develop a small peak indicating the NEC is satisfied at the location adjacent to the asymptotically flat region (refer to the insets in Figs.\,\ref{plot_wormsol_NEC}(a)-(b)), since the type I double throat appears in that limit of $\beta$. Nevertheless, the peak disappears for large values of $\lambda$ because type I double throat doesn't exist anymore. Similarly, for the case of $\lambda=30$ and $70$, the violation of scaled NEC at the throat $x=0$ decreases when $\beta$ increases from zero, we see that the scaled NEC is satisfied at $x=0$ when the type II double throat configuration appears with the throat at $x=0$ becomes the equator. As $\beta$ continues to increase and type II double throat configuration starts to diminish, the scaled NEC decreases and becomes negative, hence it is violated again when the wormholes become single throat again until $\beta$ approaches the critical value. Figs.\,\ref{plot_wormsol_NEC}(e)-(f) show the NEC of $\lambda$=2 for type I + II double throat configuration which is related to Fig.\,\ref{plot_geom_throat_1_2}(a). Since the wormholes possess two types of double throat, they certainly exhibit the phenomena of NEC due to these double throats. When the wormholes possess a type II double throat in the range of $0.8806 \leq \beta \leq 1.0074$ (left inset of Fig.\,\ref{plot_geom_throat_1_2}(a)), the NEC at $x=0$ assumes a small positive value which is slightly above the zero (inset of Fig.\,\ref{plot_wormsol_NEC}(e)). Similarly, a small peak of positive NEC is noticed (inset of Fig.\,\ref{plot_wormsol_NEC}(f)) when the wormholes exhibit a type I double throat near the asymptotically flat region (right inset of Fig.\,\ref{plot_geom_throat_1_2}(a)). However, in the BPS limit \cite{Chew:2020svi}, the violation of NEC decreases with the increase of $\beta$.

\subsection{Junction Condition}

We follow the approach of Kanti et. al. \cite{Kanti:2011yv} to evaluate the Einstein-matter field equations at $\eta=0$ for studying the junction condition of wormholes,
\begin{equation}
 \left< G^\mu\,_\nu-\beta T^\mu\,_\nu \right>=s^\mu\,_\nu \,, \quad  \left< D_\mu F^{\mu \nu} - \frac{i}{4} \left[  \Phi, D^\nu \Phi  \right] \right> = \tilde{s}^\nu \,, \quad  \left< D_\mu D^\mu \Phi - \lambda \left(   \Phi^2 - \upsilon^2  \right) \Phi \right> = s_2 \,.
\end{equation}
Here we have denoted by $s^\mu\,_\nu$ the stress-energy tensor for the matter at the throat, by $\tilde{s}^\nu$ the source term for the SU(2) vector fields and by $s_2$ the source term for the Higgs field. To obtain the left-hand side of the junction conditions, one has to integrate the system of EYMH equations across the boundary $l=0$, i.e., to evaluate the expressions 
\begin{align}
 \left< G^\mu\,_\nu-\beta T^\mu\,_\nu \right> &= \frac{1}{2} \lim_{L\rightarrow 0}  \int_{-L}^{L} \left(  G^\mu\,_\nu-\beta T^\mu\,_\nu   \right) dl \,,  \\
 \left< D_\mu F^{\mu \nu} - \frac{i}{4} \left[  \Phi, D^\nu \Phi  \right] \right> &=  \frac{1}{2}  \lim_{L\rightarrow 0}   \int_{-L}^{L} \left(  D_\mu F^{\mu \nu} - \frac{i}{4} \left[  \Phi, D^\nu \Phi  \right]   \right) dl \,,  \\
 \left< D_\mu D^\mu \Phi - \lambda \left(   \Phi^2 - \upsilon^2  \right) \Phi \right> &= \frac{1}{2}  \lim_{L\rightarrow 0}  \int_{-L}^{L} \left(  D_\mu D^\mu \Phi - \lambda \left(   \Phi^2 - \upsilon^2  \right)   \right) dl \,,
\end{align}
where $dl=\sqrt{F_1} d\eta$.

Hence, the EYMH equations read
\begin{align}
  G^t\,_t-\beta T^t\,_t &=   \frac{F''_1}{F^2_1} + \frac{2 \eta}{h F^2_1} F'_1 -  \frac{3 F'^2_1}{4 F^3_1} +  \frac{\eta_0^2 F_1 }{h^2 F_1}  - \beta \frac{1}{2 F_1} \psi'^2 + \beta \frac{(K^2-1)^2 + 2 h K'^2 + h^2 F_1 H'^2 + 2 h F_1 H^2 K^2}{ 2 F^2_1 h^2 }  \nonumber \\
&  \quad  + \beta \frac{\lambda}{4} \left( H^2-\upsilon^2 \right)^2     \,, \label{ode1} \\
 G^\eta\,_\eta-\beta T^\eta\,_\eta &=   \left(  \frac{F'_1}{2 F_1} + \frac{\eta}{h} \right) \frac{F'_0}{F_0 F_1}   + \frac{F'^2_1}{4 F^3_1} +  \frac{\eta}{h F^2_1} F'_1 -   \frac{\eta_0^2}{h^2 F_1}  +   \frac{ \beta}{2 F_1} \psi'^2  \nonumber \\
& \quad - \beta  \frac{ - (K^2-1)^2 + 2 h K'^2 + h^2 F_1 H'^2 - 2 h F_1 H^2 K^2}{2 h^2 F^2_1}  + \beta \frac{\lambda}{4} F_1 (H^2-\upsilon^2)^2  \,,  \label{ode2}  \\
 G^\theta\,_\theta-\beta T^\theta\,_\theta &=  \frac{F''_0}{2 F_0 F_1} + \left( - \frac{F'_0}{4 F^2_0 F_1} + \frac{\eta}{2 h F_0 F_1} \right) F'_0 +  \frac{1}{2 F^2_1} F''_1 +  \left( - \frac{F'_1}{F_1} + \frac{\eta}{h} \right) \frac{F'_1}{2 F^2_1}   + \frac{ \eta_0^2}{ h^2 F_1}  - \beta \frac{\psi'^2}{2 F_1} \nonumber \\
& \quad +   \frac{-(K^2-1)^2 + h^2 F_1 H'^2}{ 2 h^2 F^2_1}  + \beta \frac{\lambda}{4}   (H^2-\upsilon^2)^2   \,, \label{ode3}    \, \\
 G^\phi\,_\phi-\beta T^\phi\,_\phi &= \sin^2 \theta  \left( G^\theta\,_\theta-\beta T^\theta\,_\theta \right) \,, \\
D_\mu F^{\mu \theta} - \frac{i}{4} \left[  \Phi, D^\theta \Phi  \right] &= \left[  - \frac{K''}{2 h F^2_1}  - \frac{1}{4 h F^2_1} \left(  \frac{F'_0}{F_0} - \frac{F'_1}{F_1}    \right) K' + \frac{K (K^2-1+ h F_1 H^2)}{2 h^2 F^2_1}  \right] \tau_\phi  \,,  \label{odeYM1}  \\
D_\mu F^{\mu \phi} - \frac{i}{4} \left[  \Phi, D^\phi \Phi  \right] &= -  \left[  - \frac{K''}{2 h F^2_1}  - \frac{1}{4 h F^2_1} \left(  \frac{F'_0}{F_0} - \frac{F'_1}{F_1}    \right) K' + \frac{K (K^2-1+ h F_1 H^2)}{2 h^2 F^2_1}  \right]  \frac{\tau_\theta}{\sin \theta}  \,,    \label{odeYM2}  \\
D_\mu D^\mu \Phi - \lambda \left(   \Phi^2 - \upsilon^2  \right) \Phi  &=  \left[  \frac{H''}{F_1} + \frac{1}{2 F_1} \left(  \frac{F'_0}{F_0} + \frac{F'_1}{F_1} + \frac{4 \eta}{h}   \right) H' - \frac{2 K^2}{h F_1} H - \lambda  ( H^2-\upsilon^2  ) H \right] \tau_\eta \,. \label{odeHiggs} 
\end{align}

Next, we evaluate the left-hand sides of the above contributions
\begin{align}
  \left<  G^t\,_t-\beta T^t\,_t  \right> &= \frac{F'_1}{F_1}  \,,  \\
 \left< G^\eta\,_\eta-\beta T^\eta\,_\eta  \right> &=  0 \,, \\
 \left< G^\theta\,_\theta-\beta T^\theta\,_\theta  \right> &=  \frac{F'_0}{2 F_0} +  \frac{F'_1}{2 F_1}  \,, \\
 \left< G^\phi\,_\phi-\beta T^\phi\,_\phi  \right> &= \sin^2 \theta   \left<  \left( G^\theta\,_\theta-\beta T^\theta\,_\theta \right) \right> \,, \\
 \left< D_\mu F^{\mu \theta} - \frac{i}{4} \left[  \Phi, D^\theta \Phi  \right]  \right> &=  - \frac{K'}{2 h F_1} \tau_\phi \,, \\
 \left< D_\mu F^{\mu \phi} - \frac{i}{4} \left[  \Phi, D^\phi \Phi  \right]  \right> &= \frac{K'}{2 h F_1} \frac{\tau_\theta}{\sin \theta}  \,,  \\
  \left< D_\mu D^\mu \Phi - \lambda \left(   \Phi^2 - \upsilon^2  \right) \Phi  \right> &= H' \tau_\eta \,,
\end{align}
where the right-hand sides should be evaluated at the throat $\eta=0$. Therefore we need to take into account the expansion at $\eta=0$ by assuming the following forms,
\begin{align}
 F_0 (\eta) &= F_{00} + F_{02} \eta^2 + F_{03} \eta^3 + F_{04} \eta^4 + ...... \,, \\
 F_1 (\eta) &= F_{10} + F_{12} \eta^2 + F_{13} \eta^3 + F_{14} \eta^4 + ...... \,, \\
 K(\eta) &=   1 + K_1 \eta +  K_{2} \eta^2 + K_{3} \eta^3 + K_{4} \eta^4 + ...... \,, \\
 H (\eta) &=  H_1 \eta + H_{2} \eta^2 + H_{3} \eta^3 + H_{4} \eta^4 + ...... \,.
\end{align}
 We substitute them into Eqs. \eqref{Minset1}-\eqref{Minset4}, some of the leading terms are given by 
\begin{align}
 F_{02} &=  \frac{\beta F_{00} \left(  4 K^2_1 - \lambda \eta^2_0 F^2_{10} \upsilon^4  \right)}{4 \eta^2_0 F_{10}  }     \,, \\
 F_{12} &=   - \frac{1}{4} \beta \lambda F^2_{10} \upsilon^4   \,, \\
K_2 &= 0 \,, \\
H_2 &= 0 \,, \\
F_{03} &= 0 \,, \\
F_{13} &= 0 \,, \\
K_3 &= - \frac{ K_1 \left( \beta K^2_1 -2 F_{10}   \right)  }{6 \eta^2_0 F_{10}}    \,, \\
H_3 &= - \frac{H_1 \left( 2 \beta K^2_1 - \beta \lambda \eta^2_0 \upsilon^4 F^2_{10} + 2 \eta^2_0 \lambda \upsilon^2 F^2_{10}  \right)}{12 \eta^2_0  F_{10}}     \,, \\
F_{04} &= \frac{ \beta K^2_1 F_{00}  \left( F_{10} + \beta K^2_1  \right)  }{6 \eta^4_0 F^2_{10}}  - \frac{\beta \lambda \upsilon^2 F_{00} \left( \beta \upsilon^2  K^2_1  - F_{10} \upsilon^2 - \eta^2_0 H^2_1 F_{10}  \right)}{12 \eta^2_0} + \frac{\beta^2 \lambda^2 \upsilon^8  F_{00} F^2_{10}}{48} \,, \\
F_{14} &= -\frac{ \beta \left( 3 K^2_1 + \eta^2_0 H^2_1 F_{10}  \right)  }{6 \eta^4_0}  + \frac{\beta \lambda \upsilon^2  F_{10} \left( 4 \upsilon^2 F_{10}  + 2 \eta^2_0 H^2_1 F_{10} +  \beta \upsilon^2  K^2_1 \right)}{24 \eta^2_0 } + \frac{ \beta^2 \lambda^2 \upsilon^8 F^3_{10}}{48}  \,, \\
K_4 &= \frac{\eta^2_0 H^2_1 F_{10}+ 3 K^2_1}{12 \eta^2_0 } \,, \\
H_4 &= \frac{H_1 K_1}{3 \eta^2_0} \,.
\end{align}

 The expansion shows that $F'_0(0)=F'_1(0)=0$, whereas $K'_0(0)$ and $H'(0)$ are finite. Therefore we only need to include the source terms for the vector field and for the Higgs field. Thus we assume
\begin{equation}
 \tilde{s}^\nu = g^{\nu \mu} \tilde{s}_\mu \,, \quad \tilde{s}_\mu dx^\mu = \frac{\upsilon}{2} \left( \tau_\phi d\theta - \tau_\theta \sin \theta d \phi     \right) \,, \quad s_2 = s \tau_\eta \,.
\end{equation}

The two vector equations then yield 
\begin{equation}
 - K' = \upsilon \,,
\end{equation}
while the Higgs equation yields 
\begin{equation}
 H' = s \,,
\end{equation}
with the left-hand sides evaluated at the throat.

\section{Conclusion and Outlook}

We have constructed the symmetric wormholes in which the throat is supported by the phantom field in the Einstein-Yang-Mills-Higgs (EYMH) system beyond the Bogomol'nyi-Prasad-Sommerfield (BPS) limit by including the Higgs potential with the Higgs self-interaction value $\lambda$. The wormhole spacetime is symmetric with respect to the radial coordinate $\eta=0$. Analogous to the BPS limit \cite{Chew:2020svi}, the wormholes possess the probe limit which is the Yang-Mills-Higgs (YMH) field in the background of the Ellis wormhole when the gravity is absent. In the presence of gravity, the wormholes possess the non-trivial non-abelian hair where the families of hairy wormholes solutions emerge from the Ellis wormhole. In the BPS limit, the mass of wormholes and the scalar charge of the phantom field increase monotonically when the gravitational strength increases. However, beyond the BPS limit by increasing $\lambda$, the masses of wormholes initially increase from null to a maximum positive value but then decrease dramatically to a negative value when the gravitational strength approaches a critical value for small values of $\lambda$ but strictly decreases to negative for high values of $\lambda$. The scaled scalar charge for the phantom field also increases very steeply at the critical value of gravitational strength compare to the BPS case. Note that when $\lambda$ increases, the domain of existence of wormholes decreases, thus the drastic changing of properties of wormholes occurs earlier since the gravitational strength approaches the critical value earlier. This is in contrast to the BPS limit where it can take very large values of gravitational strength \cite{Chew:2020svi}.

The gauge field and Higgs field behave monotonically where the gauge field assumes a maximum value at $\eta=0$ and decreases to zero at the asymptotic region while the Higgs field increases from zero at $\eta=0$ to its asymptotic value. Beyond the BPS limit when the gravitational strength approaches a critical value, the Higgs field increases very steeply toward its asymptotic value while the gauge field approaches unity. Similarly, the behaviour of metric functions also changes very sharply toward their asymptotic values at infinity. Furthermore, the precision of wormhole solutions drop significantly when the gravitational strength approaches the critical value, thus we are not able to study the properties of wormholes beyond that critical value.

The hairy wormholes possess a double throat configuration in the BPS limit when the gravitational strength exceeds a value. By varying $\lambda$, the hairy wormholes exhibit richer geometrical structures than their counterpart in the BPS limit for the double throat configurations, namely, type I, type II and the hybrid of type I and II. Type I and II double throat configurations exist for $0 <\lambda < 3.0283$ and $1.8814 \leq \lambda \leq 100$, respectively. The hybrid of type I and II appears for $1.8814 \leq \lambda <  3.0283$ where type I and II configurations can coexist together. 

In type I the wormholes only exhibit double throat configuration when the gravitational strength approaches a critical value where the throat at $\eta=0$ still remains but simultaneous another throat is formed near the asymptotically flat region and the equator is sandwiched between them. However, the wormholes merely possess a single throat when the gravitational strength before approaching the critical value. 

On the contrary, in type II double throat configuration the wormholes start to develop the double throat configuration for small values of gravitational strength which is far below the critical value. The throat at $\eta=0$ becomes the equator and another throat concurrently develops in the vicinity of $\eta=0$. As gravitational strength further increases, the distance between that throat and the equator increases to a maximum and then decreases until that throat disappears. Hence, the wormholes transform back from double throat to single throat until the gravitational strength approaches the critical value. Besides, we find that type II double throat appears earlier for small gravitational strength when $\lambda$ increases. Moreover, the hybrid of type I and II double throat configuration acts as a transition state from type I to II, where type I gradually disappears with the increase of $\lambda$. Therefore, the non-zero of Higgs self-interaction value can give rise to the interesting geometrical structure of double throat for the wormholes.

The null energy condition (NEC) is most violated at the throat $\eta=0$ when the gravitational strength vanishes since the Ellis wormhole is the trivial solution. For small values of $\lambda$, the increase of gravitational strength initially weakens the violation of NEC at $\eta=0$ to a minimum value which is close to zero but then the corresponding violation increases again and the NEC becomes most violated when the gravitational strength approaches the critical value. We find that the appearance of the double throat can affect the NEC. When the type I double throat exists, a small peak of NEC develops near the asymptotically flat region when the gravitational strength approaches the critical value, hence the NEC is satisfied there. However, the peak disappears for large value of $\lambda$ because type I double throat doesn't exist anymore. Whereas for those wormholes with a higher value of $\lambda$, as the gravitational strength increases, the violation of NEC at $\eta=0$ is also weakening and then NEC can become positive, thus NEC can be satisfied also with the appearance of type II double throat configuration. However, when type II double throat gradually disappears, the NEC at $\eta=0$ decreases and then becomes negative, hence it is most violated when the gravitational strength approaches the critical value. When the wormholes possess the hybrid of type I and II double throats, they indeed exhibit the phenomena of violation of NEC for small and large values of $\lambda$. In the BPS limit, the violation of NEC decreases when the gravitational strength increases.

On the other hand, since the first-order derivatives of YMH fields are finite at $\eta=0$, thus there will be a discontinuity of YMH equations which are evaluated at $\eta=0$, therefore we need to introduce the extra source terms to overcome the discontinuity.

Let us comment on the stability issue of the non-Abelian wormholes. As was pointed out in our previous paper \cite{Chew:2020svi}, the wormholes supported by a phantom field are generally unstable against linear perturbations \cite{Gonzalez:2008wd,Gonzalez:2008xk,Torii:2013xba,Dzhunushaliev:2013lna,Dzhunushaliev:2014mza,Aringazin:2014rva}. Likewise, the particle-like and hairy black hole solutions of the EYMH system are unstable as well \cite{greene1993eluding,winstanley1995instability,mavromatos1996aspects}. Therefore, we conjecture that the EYMH hairy wormholes will inherit these instabilities and should behave qualitatively unstable. Hence, in the future, we plan to carry out an exhaustive numerical analysis of full linear stability for hairy wormholes consistently by perturbing all the functions. However, the calculation of (un)stable modes would be tedious and very challenging since the presence of the YMH field makes the calculation becomes non-trivial. Besides, the unstable modes disappear for sufficiently rapidly rotating 5-dimensional Ellis wormholes with equal angular momenta \cite{Dzhunushaliev:2013jja}. Since the counterpart EYMH black holes can rotate, then it is interesting to construct the rotating EYMH wormholes which might be stable against the perturbation. 

Moreover, the static and regular EYMH solutions can also possess only axial symmetry and need not be spherically symmetric, their counterpart static black holes also can possess only axially symmetric horizon \cite{Hartmann:2000gx,Hartmann:2001ic} which are the counterexamples to Israel's theorem. Therefore, as a first step to constructing the rotating wormholes in EYMH, we could consider constructing the static hairy wormhole solutions with a throat that is also axially symmetric.

\section*{Acknowledgement}
XYC is supported by the National Research Foundation of Korea (Grant No.:  2021R1C1C1008622, 2021R1A4A5031460). We really appreciate having a useful discussion with Jutta Kunz and Burkhard Kleihaus about the junction condition of wormhole.

\bibliography{worm}

\end{document}